\documentclass[letterpaper,12pt,twoside]{article}
\usepackage[latin1]{inputenc}
\usepackage{times}
\usepackage{epsfig}
\usepackage{dsfont}
\usepackage{amsfonts}
\usepackage{float}
\usepackage{amsmath}
\usepackage{latexsym}
\usepackage{xr} 
\usepackage{natbib}
\usepackage{color}
\usepackage{multirow}
\usepackage{setspace}

\relax 
\allowdisplaybreaks[3]

\newtheorem{definition}{{\sc Definition}\sc}[section]
\newcommand{\bdefi}{\begin{definition}}
\newcommand{\edefi}{\end{definition}}

\newtheorem{appropr}[definition]{{\sc Approximation Procedure}\sc}
\newcommand{\bappr}{\begin{appropr}}
\newcommand{\eappr}{\end{appropr}}

\newtheorem{bedi}[definition]{{\sc Condition}\sc}
\newcommand{\bbd}{\begin{bedi}}
\newcommand{\ebd}{\end{bedi}}

\newtheorem{bedin}[definition]{{\sc Conditions}\sc}
\newcommand{\bbdn}{\begin{bedin}}
\newcommand{\ebdn}{\end{bedin}}

\newtheorem{corollary}[definition]{{\sc Corollary}\sc}
\newcommand{\bco}{\begin{corollary}}
\newcommand{\eco}{\end{corollary}}

\newtheorem{lemma}[definition]{{\sc Lemma}\sc}
\newcommand{\blem}{\begin{lemma}}
\newcommand{\elem}{\end{lemma}}

\newtheorem{proposition}[definition]{{\sc Proposition}\sc}
\newcommand{\bpro}{\begin{proposition}}
\newcommand{\epro}{\end{proposition}}

\newtheorem{satz}[definition]{{\sc Theorem}\sc}
\newcommand{\bsa}{\begin{satz}}
\newcommand{\esa}{\end{satz}}

\newtheorem{theorem}[definition]{{\sc Theorem}\sc}
\newcommand{\bth}{\begin{theorem}}
\newcommand{\eth}{\end{theorem}}

\newtheorem{assumption}[definition]{{\sc Assumption}\sc}
\newcommand{\bas}{\begin{assumption}}
\newcommand{\eas}{\end{assumption}}

\newtheorem{assumptions}[definition]{{\sc Assumptions}\sc}
\newcommand{\bass}{\begin{assumptions}}
\newcommand{\eass}{\end{assumptions}}

\newtheorem{abb}{{\sc Figure}\sc}
\newcommand{\babb}{\begin{abb}}
\newcommand{\eabb}{\end{abb}}

\newenvironment{remark}{\begin{rmk}\sl}{\end{rmk}}
\newtheorem{rmk}{{\sc Remark}\sc}[section]
\newcommand{\brem}{\begin{remark}}
\newcommand{\erem}{\end{remark}}

\newenvironment{remarks}{\begin{rmks}\sl}{\end{rmks}}
\newtheorem{rmks}{{\sc Remarks}\sc}[section]
\newcommand{\brems}{\begin{remarks}}
\newcommand{\erems}{\end{remarks}}

\newenvironment{example}{\begin{exmp}\rm}{\end{exmp}}
\newtheorem{exmp}{{\sc Example}\sc}[section]
\newcommand{\bbsp}{\begin{example}}
\newcommand{\ebsp}{\end{example}}
\newcommand{\bexa}{\begin{example}}
\newcommand{\eexa}{\end{example}}

\newtheorem{model}{{\sc Model}\sc}[section]
\newcommand{\bmdl}{\begin{model}}
\newcommand{\emdl}{\end{model}}

\newtheorem{scheme}{{\sc Scheme}\sc}[section]
\newcommand{\bscm}{\begin{scheme}}
\newcommand{\escm}{\end{scheme}}

\newenvironment{tabelle}{\begin{tabl}\rm}{\end{tabl}}
\newtheorem{tabl}{{\bf Table}}
\newcommand{\btab}{\begin{tabelle}}
\newcommand{\etab}{\end{tabelle}}

\newenvironment{exercise}{\begin{exc}\sl}{\end{exc}}
\newtheorem{exc}{Exercise}[section]
\newcommand{\bexe}{\begin{exercise}}
\newcommand{\eexe}{\end{exercise}}

\newcommand{\proofit}{\noi {\it Proof:}\ }

\relax

\newcommand{\renu}{\mathbb{R}}
\newcommand{\natnu}{\mathbb{N}}

\newcommand{\noi}{\noindent}

\newcommand{\1}{{\mathds{1}}}

\newcommand{\qed}{\mbox{ } \hfill $\Box$\\ }

\newcommand{\bay}{\begin{array}}
\newcommand{\eay}{\end{array}}

\newcommand{\bqa}{\begin{eqnarray*}}
\newcommand{\eqa}{\end{eqnarray*}}

\newcommand{\bee}{\begin{eqnarray*}}
\newcommand{\eee}{\end{eqnarray*}}

\newcommand{\bea}{\begin{eqnarray*}}
\newcommand{\eea}{\end{eqnarray*}}

\newcommand{\bqan}{\begin{eqnarray}}
\newcommand{\eqan}{\end{eqnarray}}

\newcommand{\be}{\begin{eqnarray}}
\newcommand{\ee}{\end{eqnarray}}

\newcommand{\bit}{\begin{itemize}}
\newcommand{\eit}{\end{itemize}}

\newcommand{\ben}{\begin{enumerate}}
\newcommand{\een}{\end{enumerate}}

\newcommand{\beq}{\begin{equation}}
\newcommand{\eeq}{\end{equation}}

\newcommand{\bdes}{\begin{description}}
\newcommand{\edes}{\end{description}}

\newcommand{\btb}{\begin{tabular}}
\newcommand{\etb}{\end{tabular}}

\newcommand{\bcen}{\begin{center}}
\newcommand{\ecen}{\end{center}}

\newcommand{\bmp}{\begin{minipage}}
\newcommand{\emp}{\end{minipage}}

\newcommand{\ep}{\epsilon}
\newcommand{\lam}{\lambda}

\newcommand{\sumi}{\sum_{i=1}^}
\newcommand{\sumj}{\sum_{j=1}^}
\newcommand{\sumk}{\sum_{k=1}^}

\newcommand{\sums}{\sum_{s=1}^}
\newcommand{\sumr}{\sum_{r=1}^}

\newcommand{\ks}{\oplus}
\newcommand{\kp}{\otimes}
\newcommand{\bks}{\bigoplus}

\newcommand{\Cov}{\operatorname{{\it Cov}}}
\newcommand{\var}{\operatorname{{\it Var}}}
\newcommand{\Var}{\operatorname{{\it Var}}}

\newcommand{\tr}{\operatorname{tr}}

\newcommand{\diag}{\operatorname{\it diag}}

\newcommand{\rank}{\operatorname{\it rank}}

\newcommand{\vlambda}{\mbox{\boldmath $\lambda$}}

\newcommand{\olc}{\overline{c}}
\newcommand{\old}{\overline{d}}

\newcommand{\olY}{\overline{Y}}
\newcommand{\olZ}{\overline{Z}}

\newcommand{\olmu}{\overline{\mu}}

\newcommand{\vc}{\boldsymbol{c}}

\newcommand{\vn}{\boldsymbol{n}}

\newcommand{\vD}{\boldsymbol{D}}

\newcommand{\vH}{\boldsymbol{H}}
\newcommand{\vI}{\boldsymbol{I}}
\newcommand{\vJ}{\boldsymbol{J}}

\newcommand{\vP}{\boldsymbol{P}}

\newcommand{\vT}{\boldsymbol{T}}

\newcommand{\vV}{\boldsymbol{V}}

\newcommand{\vY}{\boldsymbol{Y}}
\newcommand{\vZ}{\boldsymbol{Z}}

\newcommand{\vGamma}{\boldsymbol{\Gamma}}

\newcommand{\vep}{\boldsymbol{\epsilon}}

\newcommand{\vlam}{\boldsymbol{\lambda}}

\newcommand{\vmu}{\boldsymbol{\mu}}
\newcommand{\vnu}{\boldsymbol{\nu}}

\newcommand{\vSigma}{\boldsymbol{\Sigma}}

\newcommand{\veins}{{\bf 1}}
\newcommand{\vnull}{{\bf 0}}

\newcommand{\volY}{\boldsymbol{\overline{Y}}}

\newcommand{\vwhV}{\boldsymbol{\widehat{V}}}

\newcommand{\vwhSigma}{\boldsymbol{\widehat{\Sigma}}}

\newcommand{\whnu}{\widehat{\nu}}

\newcommand{\wtQ}{\widetilde{Q}}

\newcommand{\wtW}{\widetilde{W}}

\newcommand{\wtep}{\widetilde{\epsilon}}

\numberwithin{equation}{section}

\let\proglang=\textsf
\newcommand{\pkg}[1]{{\fontseries{b}\selectfont #1}}

\begin{document}

%\doublespacing

% tit.tex
% Valid Permutation Procedures For Longitudinal Data In Semiparametric Models

\title{\Large \bf Permuting Longitudinal Data Despite All The Dependencies %- astonishingly successful
}
\author{Sarah Friedrich$^{*}$,  Edgar Brunner$^{**}$ and  Markus Pauly$^{*}$ \\[1ex] 
%{\small University of Gï¿½ttingen, Germany}
}
\maketitle

\begin{abstract}
For general repeated measures designs the Wald-type statistic (WTS) is an asymptotically valid procedure allowing for unequal covariance matrices and possibly non-normal multivariate observations. 
The drawback of this procedure is the poor performance for small to moderate samples, i.e., decisions based on the WTS may become quite liberal. It is the aim of the present paper to improve its small sample behavior by means of a novel permutation procedure. 
In particular, it is shown that a permutation version of the WTS inherits its good large sample properties while yielding a very accurate finite sample control of the type-I error as shown in extensive simulations. 
Moreover, the new permutation method is motivated by a practical data set of a split plot design with a factorial structure on the repeated measures. 
\end{abstract}


\vfill
\vfill

\noindent${}^{*}$ {University of Ulm, Institute of Statistics, Germany\\
 \mbox{ }\hspace{1 ex}email: sarah.friedrich@uni-ulm.de}

\noindent${}^{**}$ {University Medical Center G\"ottingen, Institute of Medical Statistics, Germany}

\newpage

% int.tex

\section{Motivation and Introduction}\label{int}

%$\valpha, \vbeta, \vgamma, \vGamma, \vLambda, \vmu,\vnu$ \\
%$\alpha, \beta, \gamma, \Gamma, \Lambda, \mu,\nu$ \\

In many experiments in the life, social or psychological sciences the experimental units (e.g. subjects) are repeatedly 
observed at different occasions (e.g.~at different time points) or under different treatment conditions. 
This leads to certain dependencies between observations from the same unit and results in a more complicated statistical analysis of such studies. 
In the context of experimental designs, the repeated measures are considered as levels of the \emph{sub-plot} factor. 
If several groups are observed, these are considered as levels of the \emph{whole-plot} factor. 
Typical questions in this setting concern the investigation of a group effect, 
a non-constant effect of time or different time profiles in the groups. 
Classical repeated measures models, where hypotheses are tested with Hotelling's $T^2$ \citep{Hotelling:1931} 
or Wilks's $\Lambda$ \citep{Wilks}, 
assume normally distributed observation vectors and a common covariance matrix for all groups, see e.g.~the monograph of \citet{Davis}.
In medical and biological research, however, the assumptions of equal covariance matrices and multivariate 
normally distributed outcomes are often not met and a violation of them may inflate the type-I error rates, 
see the comments in \citet{Xu:2008}, \citet{Suo:2013} or \citet{Kon:2015}. 
Therefore, other procedures have been developed for repeated measures which are based on certain approximation techniques 
(\cite{GG1958}; \cite{GG1959}; \cite{Huynh1976}; \cite{Lecoutre1991}; \cite{Keselman}; \cite{Werner:2004};
\cite{Ahmad_etal_2008}; \cite{Brunner2009}; \cite{Brunneretal2009}; \cite{Kenward}; \cite{Brunner2012}; \cite{Chi2012}; \cite{PEB}). 
However, these papers mainly assume the multivariate normal distribution and only discuss methods for 
specific models which are also asymptotically only approximations, i.e., they do not even lead to 
asymptotic exact tests. %in general not  as the sample size increases. 
Another possibility is to apply a specific mixed model in the GEE context, see e.g. the text books by Verbeke and Molenberghs (2009, 2012). 
These methods require that the data stems from a specific exponential family. 
An exception is given by the multivariate Wald-type test statistic (WTS)
%Especially for ordinal data, the normality assumption is obviously violated and nonparametric methods are needed, see %e.g. \cite{Hollander}. 
%Even though these methods do not assume normality, 
%they usually still assume that the shapes of the distributions are the same for all groups, implying equal variances.
%On the other hand, several approximations for heteroscedastic designs have been proposed, e.g.~the generalized Welch-James test by \citet{Johansen:1980} the ANOVA-type test statistic in \cite{BrDeMu:1997}, see also \eqref{ATS} below. 
which is asymptotically exact. However, it is well known that it requires large sample sizes to keep the pre-assigned 
type-I error level, see e.g. \citet{brunner:2001}, \citet{PBK} and \citet{Kon:2015}.\\
To improve the small sample behavior of the WTS in a MANOVA setting, \citet{Kon:2015} proposed different bootstrap techniques. 
Another possibility would be to apply permutation procedures. 
It is well known that permutation tests are finitely exact under the assumption of exchangeability, 
see e.g. \cite{Brombin:2013}, \cite{pesarin-book}, \cite{mielke-berry-2007} or \cite{pesarin-salmaso-book, Pesarin:2010, Pesarin:2012} for examples. 
In most of these examples, however, permutation tests are only applied in situations where the null distribution is invariant under 
the corresponding randomization group. A modified permutation procedure may also be applied in 
situations where this invariance does not hold, see e.g. \cite{janssen:2003}, \cite{janssen:2005}, \cite{OmPa:2012}, \cite{Chung:2013} and \cite{PBK}.
The main idea in these papers is to apply a studentized test statistic and to use its permutation distribution (based on permuting the pooled sample) for calculating critical values. 
This leads to particularly good finite sample properties even in case of general factorial designs with fixed factors, see \cite{PBK}.
It is the aim of the present paper to
extend the concept of permuting all data to the context of longitudinal data in general 
(not necessarily normal and homoscedastic) split plot designs. Applied to the WTS this generalizes the results of \cite{PBK} and leads 
to astonishingly accurate results despite the dependencies in repeated measurements data.\\

%The current paper has a similar aim for repeated measures settings. However, instead of employing bootstrap techniques we 

% improve the small sample behavior of the WTS in 
%the context of repeated measures by applying bootstrap and permutation procedures.
%A bootstrap approach in repeated measures designs has been proposed in , who applied robust estimation and bootstrapping methods to the Huynh Improved General Approximation (IGA) test \citep{Huynh}. Several simulation studies analyzing the behavior of the different approaches are presented in \cite{Keselman}. However, proofs for the consistency of the bootstrap approach are not provided.\\

The methodology derived in this paper is motivated by the following data example on the $O_2$ consumption of leukocytes.
To examine the breathability of leukocytes, an experiment with 44 HSD-rats was conducted. A group of 22 rats was treated with a placebo, while the other 22 rats were treated with a substance supposed to enhance the humoral immunity. 18 hours prior to the opening of the abdominal cavity, all animals received 2.4 g sodium-caseinat 
for the production of a peritoneal exudate rich on leukocytes. 
In order to obtain a sufficient amount of material the peritoneal liquid of 3--4 animals was mixed 
and the leukocytes therein were rehashed in an experimental batch. 
One half of the experimental batch was mixed with inactivated staphylococci in a ratio of 100:1, 
the other half remained untreated and served as a control. Then, the oxygen consumption of the leukocytes was measured with a polarographic 
electrode after 6, 12 and 18 minutes, respectively. For each group separately, 
12 experimental batches were carried out. The means over the experimental batches in both treatment groups are listed in Table~\ref{bsp.leuko2.tab}.
\text{ } \\[-8mm]
\begin{table}[H] 
			\centering
			\caption{\it Mean oxygen consumption of leukocytes in the presence and absence of inactivated staphylococci.}
			\label{bsp.leuko2.tab} 
			{\small
				\begin{tabular}{|c||rrr|rrr|} \hline
					\multicolumn{7}{|c|}{Mean $O_2$-Consumption [$\mu \ell$]} \\ \hline
					& \multicolumn{6}{c|}{Staphylococci} \\
					& \multicolumn{3}{c|}{With} & \multicolumn{3}{c|}{Without} \\ \hline
					& \multicolumn{3}{c|}{Time [min]} & \multicolumn{3}{c|}{Time [min]} \\
					& 6 & 12 & 18 & 6 & 12 & 18  \\ \hline \hline
					Placebo & 1.618 & 2.434 & 3.527 & 1.322 & 2.430& 3.425 \\[-1mm] \hline
					Verum & 1.656 & 2.799 & 4.029 & 1.394 & 2.57 & 3.677 \\[-1mm]
					\hline
					\hline
				\end{tabular}
			}
\end{table}

%{\color{red}Aufgabe fuer Sarah: 2 kleine  Kovarianzmatrizen angeben Placebovs. Verum }

Questions of interest in this example concern the effect of the whole-plot factor 'treatment', the effect of the sub-plot factors 'staphylococci' and 'time' as well as interactions between these effects. 
We note that the empirical $6\times 6$ covariance matrices of the two groups appear to be quite different (see the supplement for details). This also motivates to include unequal covariance matrices in our model. 
For such experimental designs procedures are derived in this paper that lead to good small sample 
control of the type-I error while being asymptotically exact. 

The paper is organized as follows. The underlying statistical model is described in Section~\ref{mod}, where we also introduce
the Wald-type (WTS) as well as the ANOVA-type statistic (ATS) and state their asymptotic behavior. 
In Section~\ref{sec:apr}, we describe the novel permutation procedure used to improve the small sample behavior of the WTS. 
Afterwards, we present the results of extensive simulation studies in Section \ref{sim}, 
analyzing the behavior of the permuted test statistic in different simulation designs with certain competitors. 
Additional simulation results have also been run for several other resampling schemes. 
They did not show a better performance than the permutation procedure and are only reported in the supplementary material, where also various power simulations can be found. 
The motivating data example is analyzed in detail in Section \ref{app}.
The paper closes with a brief discussion of our results in Section \ref{dis}. All proofs are given in the supplementary material.

 %mod.tex

\section{Statistical Model, Hypotheses and Statistics} \label{mod}

\subsection{Statistical Model and Hypotheses} 

To establish the general model, let
\bqan\label{model}
\vY_{ik}&=&(Y_{ik1}, \dots, Y_{ikt_i})', \hspace{0.5cm} i=1, \dots , a;\;\; k= 1, \dots , n_i
\eqan
denote independent random vectors with distribution $F_i$, expectation 
$\vmu_i = (\mu_{i1}, \dots, \mu_{it_i})' = E(\vY_{i1})$ and covariance matrix 
$\vV_i = \Cov(\vY_{i1}) > 0$ in treatment group $i$. % which we assume to exist. 
We do not assume \emph{any special structure} of the covariance matrix $\vV_i$ 
which may even be \emph{different between groups} $1 \leq i \leq a$. 
Note that we also allow the number of time points $t_i$ to differ between groups. The most common case where $t_i = t$ for all $i=1, \dots, a$ is thus a special case of model \eqref{model}. 
Here the time points $t_i \in \natnu$ are fixed.
For convenience, we collect the observation vectors $\vY_{ik}$ in
\bqan\label{pooled sample}
\vY &= & (\vY_1', \dots, \vY_a')', \hspace{1cm} \vY_i = (\vY_{i1}', \dots, \vY_{in_i}')'.
\eqan
In this set-up, hypotheses are formulated as $H_0^{\mu} : \vH \vmu = \boldsymbol{0}$, where $\vmu = (\vmu_1', \dots, \vmu_a')'$ denotes the 
vector of all expectations $\mu_{is} = E(Y_{i1s})$, $i = 1, \dots, a;\; s=1, \dots, t_i$ and $\vH$ is a suitable contrast matrix, 
i.e., its rows sum up to zero. Examples of $\vH$ are presented in Section~\ref{sim}.

Throughout the paper, we will use the following notation. We denote by $\vI_t$ the $t$-dimensional unit matrix and by $\vJ_t$ the $t \times t$ matrix of 1's, i.e., $\vJ_t = \boldsymbol{1}_t \boldsymbol{1}_t'$, where $\boldsymbol{1}_t=(1, \dots, 1)'$
is the $t$-dimensional column vector of 1's. Furthermore, let $\vP_t = \vI_t - \frac{1}{t} \vJ_t$ denote the $t$-dimensional centering matrix. By $\ks$ and $\kp$ we denote the direct sum and the Kronecker product, respectively.

An estimator of $\vmu$ is given by
\bqa
 \volY_{\cdot} = (\volY_{1\cdot}', \dots, \volY_{a\cdot}')' \text{ for } \volY_{i\cdot} = (Y_{i\cdot1}, \dots, Y_{i\cdot t_i})', ~ \olY_{i\cdot s} = \frac {1}{n_i} \sum_{k=1}^{n_i} {Y_{iks}}, ~ s=1, \dots, t_i
\eqa
and the covariance matrix $\vV_i$ in treatment group $i$ is estimated by the sample covariance matrix 
\bqa
\vwhV_i = \frac{1}{n_i-1} \sum_{k=1}^{n_i} (\vY_{ik}-\volY_{i \cdot}) (\vY_{ik}-\volY_{i \cdot})', ~ i=1, \dots, a.
\eqa

Let $N = \sum_{i=1}^a n_i$ denote the total number of subjects in the trial, $T = \sumi a t_i$ the total number of time points and $\tilde{N} = \sumi a n_i t_i$ the total number of observations. Then the asymptotic results are derived under the following two assumptions:
\bit
\item[(1)] $\frac{n_i}{N} \to \kappa_i \in(0,1)$ as $\min(n_1,\ldots,n_a)\to\infty$, 
\item[(2)] $\sup_{i}E(||\vY_{i1}||^4)< \infty$.
\eit

\subsection{Statistics and Asymptotics}

We consider two commonly used test statistics for repeated measures and multivariate data. First, the so-called ANOVA-type statistic (ATS), introduced in \cite{brunner:2001}, is given as:
%Another possible test statistic introduced in \cite{brunner:2001} is the so-called ANOVA-type test statistic (ATS).
%The idea is to drop the estimated covariance matrix in \eqref{WTS}. This leads to the following statistic:
\bqan \label{ATS}
\wtQ_N = N \volY_{\cdot}'\vH'(\vH \vH')^-\vH \volY_{\cdot} = N \volY_{\cdot}' \vT  \volY_{\cdot},
\eqan
where $(\cdot)^-$ denotes some generalized inverse. Note that the test statistic does not depend on the special choice of the generalized inverse.
Its asymptotic distribution is established in the next theorem.  

\begin{theorem} \label{theo:ATS}
	Under the null hypothesis $H_0^{\mu}: \vH \vmu = \boldsymbol{0}$, the ATS in \eqref{ATS} has, asymptotically, the same distribution as the random variable
	\[
	X = \sumi a \sums {t_i} \lambda_{is} X_{is},
	\]
	where $X_{is} \stackrel{i.i.d.}{\sim} \chi_1^2$ and the weights $\lambda_{is}$ are the eigenvalues of $\vT \vSigma$ for 
	$\vSigma = \bks _{i=1}^a \kappa_i^{-1} \vV_i$.\\
	Moreover, for local alternatives $\vT \vmu = \frac{1}{\sqrt{N}} \cdot \vT \vnu, \vnu \in \renu^{T}$, it holds that the ATS has, asymptotically, 
	 the same distribution as $\vZ' \vT \vZ$, where $\vZ \sim N(\vnu, \vSigma)$. If additionally $\vSigma>0$, the ATS has the same distribution as a weighted sum of $\chi^2_1(\delta)$ distributed random variables, where the weights are again the eigenvalues $\lambda_{is}$ and $\delta = \vnu' \vSigma^{-1}\vnu$.
\end{theorem}

Since the $\lambda_{is}$ are unknown, the result cannot be applied directly. Nevertheless, \cite{brunner:2001} proposed to approximate the distribution of $X$ by the 
distribution of a scaled $\chi^2$-distribution, i.e., by
$g \cdot \tilde{X}_\nu$, where $\tilde{X}_\nu \sim \chi^2_\nu$ \citep{box}. The constants $g$ and $\nu$ are estimated from the data 
such that the first two moments of $X$ and $g \cdot \tilde{X}_\nu$ coincide \citep{box}. 
This leads to approximating the statistic 
\bqan \label{ATSstatistic}
F_N = \frac{N}{\tr(\vT\vwhSigma)} \volY_{\cdot}' \vT  \volY_{\cdot}
\eqan
by an $F(\whnu, \infty)$-distribution with estimated degree of freedom 
$
\hat{\nu} = \tr^2(\vT \vwhSigma)/\tr(\vT \vwhSigma)^2$, where $\vwhSigma = N \bks _{i=1}^a \frac{1}{n_i} \vwhV_i$. % see also \cite{brunner:2001}.
The corresponding ATS test $\varphi_{ATS} = \1\{\wtQ_N > F_{\alpha}(\whnu, \infty)\}$, where $F_\alpha(\whnu, \infty)$ denotes the $(1-\alpha)$-quantile of the $F(\whnu, \infty)$-distribution, leads to consistent test decisions for fixed alternatives. However, it is in general no asymptotic level $\alpha$ test under the null hypothesis, which is a severe drawback of this procedure.
Thus, we discuss a second statistic, the so-called Wald-type statistic (WTS) given as
\bqan \label{WTS}
Q_N = N \volY_{\cdot}'\vH'(\vH \vwhSigma \vH')^+\vH \volY_{\cdot}.
\eqan
Here $(\vH \vwhSigma \vH')^+$ denotes the Moore-Penrose inverse of $(\vH \vwhSigma \vH')$. 
In order to test the general linear hypotheses $H_0^\mu: \vH \vmu = \vnull$ critical values are 
taken from the asymptotic distribution of $Q_N$ under the null hypothesis stated below.
\begin{theorem}\label{WTS_asy}
	Under the null hypothesis $H_0^{\mu}: \vH \vmu = \boldsymbol{0}$, the WTS in \eqref{WTS} has, asymptotically, a  central $\chi^2_f$-distribution with $f = \rank(\vH)$.
	The corresponding test is given by $\varphi_{WTS} = \1\{Q_N > \chi^2_{f, 1-\alpha}\}$, where $\chi^2_{f,1-\alpha}$ denotes the $(1-\alpha)$-quantile of the $\chi^2_f$ distribution. This test is an asymptotic level $\alpha$ test and is consistent for general fixed alternatives $\vH \vmu \neq \vnull$. Moreover, for local alternatives $\vH \vmu = \frac{1}{\sqrt{N}} \cdot \vnu, \vnu \in \renu^{T}$, it holds that $Q_N$ has asymptotically a non-central $\chi^2_f(\tilde{\delta})$ distribution where $\tilde{\delta}=(\vH \vnu)'(\vH \vSigma\vH')^+\vH \vnu$. This implies that $E_{H_1}(\varphi_{WTS}) \to P(Z > \chi^2_{f, 1-\alpha})$ with $Z \sim \chi^2_f(\tilde{\delta})$.
\end{theorem}
Although $\varphi_{WTS}$ possesses these nice asymptotic properties, it is well-known that very large sample sizes $n_i$ are necessary to maintain the pre-assigned level $\alpha$ using quantiles of the 
limiting $\chi^2$-distribution, see \cite{Kon:2015}, \cite{PBK} and \cite{brunner:2001} as well as Table \ref{table:1Sp_ln} below. This leads to a limited applicability of the WTS in practice.

To accept the need for a novel procedure, we investigate the accuracy of the two test statistics %approximation 
in a one sample repeated measure design with $n$ subjects and $t$ repeated measures $Y_{ks}$. 
The null hypothesis $H_0^{\mu}: \{\mu_1 = \dots = \mu_t\}= \{\vP_t \vmu = \vnull\}, \vmu=(\mu_1, \dots, \mu_t)'$ is considered and the components of $\vY_{k}$ 
are selected as standardized log-normally distributed random variables, i.e.,
\bqa
Y_{ks} = \frac{{\ep}_{ks} - E({\ep}_{ks})}{\sqrt{\Var({\ep}_{ks})}}
\eqa
for i.i.d.~log-normally distributed ${\ep}_{ks}$, $k=1, \dots, n$ and $s=1, \dots, t$. 
The results are displayed in Table \ref{table:1Sp_ln}, where the simulated type-I error rates of the WTS and ATS are given. 
It is readily seen that the test based on the WTS considerably exceeds the nominal level of 5\%, 
while the ATS leads to rather conservative decisions. 

\begin{table}[H]
	\centering
	\caption{\it Simulated type-I error rates (10000 simulations) in a repeated measures design with $n=10, 20, 50, 100$ individuals and $t=4, 8$ repeated measures. The ATS is compared to the upper 5\% quantile of the $F(\hat{\nu}, \infty)$-distribution, 
	the WTS to the upper 5\% quantile of the $\chi^2_{t-1}$-distribution.}
	\label{table:1Sp_ln}
	\vspace{0.3cm}
	\begin{tabular}{|c||c|c||c|c|}
		\hline
		& \multicolumn{4} {c|} {Type-I error rates ($\alpha=0.05$) } \\	\hline
		& \multicolumn{2}{c||}{ATS: F-quantile} & \multicolumn{2}{c|}{WTS: $\chi^2$-quantile} \\ \hline
		$n$ & $t=4$ & $t=8$ & $t=4$ & $t=8$ \\ \hline
		10	& 0.025	  &  0.012  &  0.223  & 0.776 \\ \hline
		20	&	0.026  &  0.014  &  0.126  &  0.388 \\ \hline
		50	&	0.030  &  0.021  &   0.081 &  0.166  \\ \hline
		100	&	0.035  &   0.025 &  0.067  & 0.111  \\ \hline
		
	\end{tabular}
\end{table}

Thus, to enhance the small sample properties of the above tests we have compared different resampling approaches in an extensive simulation study, 
presented in Section \ref{supp:OtherResampling} of the supplementary material \cite{Friedrich_supp}. Surprisingly, 
the best procedure turned out to be a permutation technique that randomly permutes 
the pooled univariate observations without taking into account the existing dependencies for calculating critical values. 
This at first sight counter-intuitive method is motivated from \cite{KoPa:2014}, 
where a similar approach has been applied in the paired two-sample case.
Moreover, the current procedure generalizes the permutation test on independent observations by \cite{PBK} and implemented in the \proglang{R} package \pkg{GFD} \citep{GFD} to the case of repeated measures and multivariate data. 
The details are explained in the next section. %It is presented in the following. 

\section{The Permutation Procedure}\label{sec:apr}

Let $\vY^{\pi}= \pi(Y_{111}, \cdots, Y_{an_at_a})' = 
(Y_{111}^\pi, \cdots, Y_{an_at_a}^\pi)'$ denote a fixed but arbitrary permutation of all $\tilde{N}$ elements of 
$\vY$ in \eqref{pooled sample}, i.e., $\pi \in \mathcal{S}_{\tilde{N}}$. In this notation, 
$Y_{iks}^{\pi}$ denotes the $(i, k, s)-$component of the 
permuted vector $\vY$. Furthermore, let $\volY_{\cdot}^{\pi}$ denote the vector of the means under 
this permutation and $\vwhSigma^\pi =  \bks_{i=1}^a \frac{N}{n_i} \vwhV_i^\pi$ the empirical covariance matrix of the permuted observations. 

It is obvious, that $\vY$ and $\vY^{\pi}$ only have the same distribution, if the components of $\vY$ are exchangeable. 
However, this is not the case in general two- and higher way layouts, even in the case of independent observations, see e.g.~\cite{Huang}. 
Following the approach of \cite{Neuhaus:1993}, \cite{janssen:1997, janssen:2005}, \cite{OmPa:2012}, \cite{Chung:2013}  
and \cite{PBK} in the case of independent observations, the idea is to studentize the statistic $\sqrt{N}\volY_{\cdot}^{\pi}$ and consider its projection into the hypothesis space, resulting in the WTS of the permuted observations, 
namely
\bqan\label{WTPS}
Q_N^{\pi} = N (\volY_{\cdot}^{\pi})'\vH'(\vH \vwhSigma^{\pi} \vH')^+\vH \volY_{\cdot}^{\pi}.
\eqan
In the sequel we will denote $Q_N^{\pi}$ as the WTPS. 
Note, that the question how to permute is more involved here than in the case of independent univariate observations. 
%The idea to randomly permute the pooled sample without taking into account the dependencies is motivated from \cite{KoPa:2014}.
A heuristic reason why the above approach might work is as follows: 
Unconditionally, all permuted components possess the same mean. Thus, when multiplied by a contrast matrix the permuted means vector 
always mimics the null situation, i.e., $\vH E(\volY_{\cdot}^{\pi}) =\vnull$ always holds. In particular, it can be shown
that the conditional distribution of the 
WTPS $Q_N^{\pi}$ in \eqref{WTPS} always approximates the null distribution of $Q_N$ in \eqref{WTS} in the general 
repeated measures design under study; thus leading to an asymptotically valid permutation test. % our case with certain possible dependencies. 
This result is formulated in the following theorem:

\begin{satz}\label{theo}
	The studentized permutation distribution of $Q_N^{\pi}$ in \eqref{WTPS} conditioned on the observed data $\vY$ 
	weakly converges to the central $\chi^2_f$ distribution in probability, where $f=rank(\vH)$.
\end{satz}

\brem\label{rem:1}
%\hspace{0.001cm}
Theorem \ref{theo} states that the permutation distribution asymptotically provides a valid approximation of the null distribution 
of the test statistic $Q_N$ in \eqref{WTS}. To be concrete, this means that for any underlying parameters $\vmu \in \renu^{T}$ and 
$\vmu_0 \in H_0(\vH)$ with $\vH\vmu_0=\vnull$ we have convergence in probability
\bqan \label{equivalence}
\sup_{x \in \renu} \left| P_{\vmu}(Q_N^{\pi}\leq x | \vY) - P_{\vmu_0}(Q_N \leq x) \right| \to 0.
\eqan
Here, $P_{\vmu}(Q_N \leq x)$ and $P_{\vmu}(Q_N^\pi \leq x | \vY)$ denote the unconditional and conditional distribution function of $Q_N$ 
and $Q_N^\pi$, respectively, under the assumption that $\vmu$ is the true underlying parameter.
%Property \eqref{equivalence} is a desirable property of a resampling procedure, see e.g.~Theorem 1 in \cite{DP:2009} or \cite{PBK} for similar examples.
\erem
\brem\label{rem:2}
A Wald-type permutation test is obtained by comparing the original test statistic $Q_N$ with the $(1-\alpha)$-quantile $c^*_{1-\alpha}$ of the conditional 
distribution of the WTPS $Q_N^{\pi}$ given the observed data $\vY$, i.e., $\varphi_{WTPS} = \1\{Q_N > c^*_{1-\alpha}\}$. Theorem \ref{theo} implies that this test asymptotically keeps the
pre-assigned level $\alpha$ under the null hypothesis and is consistent for any fixed alternative $\vH \vmu \neq \textbf{0}$, 
i.e., it has asymptotically power 1. Moreover, it has the same asymptotic power as the WTS for local alternatives $\vH \vmu = \frac{1}{\sqrt{N}}\cdot \vnu$, i.e., it holds that $E_{H_1}(\varphi_{WTPS}) \to P(Z > \chi^2_{f, 1- \alpha})$ with $Z \sim \chi^2_f(\tilde{\delta})$ as in Theorem \ref{WTS_asy}.

It follows that the permutation test and the classical Wald-type test are asymptotically equivalent and 
that both have the same local power under contiguous alternatives. In particular the asymptotic relative efficiency of the WTPS compared to the classical WTS is 1.
Moreover, the permutation test based on $Q_N^{\pi}$ is finitely exact if the pooled data $\vY$ are exchangeable under the null hypothesis. 
In comparison, the ATS also leads to a consistent test for fixed alternatives but does not provide an asymptotic level 
$\alpha$ test since it is only an approximation. 

We note, that the proof given in the supplement to this paper indicates that the given permutation technique does not work in case of the ATS. In particular, a 
permutation version of the ATS would also possess a weighted $\chi^2$-limit distribution but with different weights, 
say $\tilde{\lambda}_{is}$, due to an incorrect covariance structure.
\erem

\brem\label{rem:3} 
Our general framework \eqref{model} allows for the treatment of different important factorial designs in 
the context of multivariate repeated measures data analysis. 
As in \cite{PBK} the idea is to accordingly split the indices in subindices 
and to choose an appropriate hypothesis matrix $\vH$. 
Examples of different cross-classified and hierarchically nested designs are discussed in Section~4 of \cite{Kon:2015}. 
For repeated measures, examples are given in Sections~\ref{sim} and \ref{app} below as well as in \cite{brunner:2001}.
\erem
%This also shows that a permutation approach in the context of longitudinal data only works with the 'right' studentized test statistic, otherwise this approach leads to wrong results.

% sim.tex 
\section{Simulations} \label{sim}
In order to investigate the small sample behavior of the WTPS, we present extensive simulation results for 
different designs and covariance structures. The procedure is analyzed in different settings with regard to maintaining 
the pre-assigned type-I error rate ($\alpha=5\%$). The results for the WTPS are compared to the asymptotic 
quantiles of the ATS ($F$-quantile) and the WTS ($\chi^2$-quantile).

\subsection{Data Generation}
For our simulation studies, we simulated a split plot design which, in the context of longitudinal data, is a design with $a$ groups, $n_i$ subjects in group $i$ and $t_i = t$ repeated measures $Y_{iks}, s=1, \dots, t$. 
Let 
\bqa
\vY_{ik} = (Y_{ik1},\ldots,Y_{ikt})' = \vmu_i + B_{ik}\veins_t + \vV_i^{1/2} \vep_{ik},
\eqa
with $\vmu_i=E(\vY_{i1}), i=1,\dots,a,$ and let $B_{ik} \sim N(0, \sigma_i^2)$ denote independent additive subject effects. 
The i.i.d.~random vectors $\vep_{ik} = (\ep_{ik1}, \dots, \ep_{ikt})$ were generated from different standardized distributions by
\bqa
\ep_{iks} = \frac{\tilde{\ep}_{iks} - E(\tilde{\ep}_{iks})}{\sqrt{\Var(\tilde{\ep}_{iks})}},
\eqa
where $\tilde{\ep}_{iks}$ denote i.i.d.~normal, exponential or log-normal random variables. 
% and the others are not shown here.\\
%(\mu_{i1}, \dots, \mu_{it}) = 0

A simulation setting with $a=3$ groups and $t=4, 8$ repeated measures was considered. 
The null hypotheses investigated are 

\ben
\item[(1)] The hypothesis of \emph{no time effect $T$}
\bqa
H_0^{\mu}(T) &:&  \olmu_{ \cdot 1}= \dots= \olmu_{\cdot t} \quad\text{or equivalently}\quad  \vH_T\vmu = \boldsymbol{0},
\eqa
\item[(2)] The hypothesis of \emph{no group $\times$ time interaction effect $GT$}
\bqa
H_0^{\mu}(GT) &:&  \vH_{GT}\vmu = \left(
\begin{array}{l}
	\mu_{11} - \olmu_{1 \cdot} - \olmu_{\cdot 1} + \olmu_{\cdot \cdot}  \\
	\vdots  \\
	\mu_{at} - \olmu_{a \cdot} - \olmu_{\cdot t} + \olmu_{\cdot \cdot} 
\end{array} \right) = \boldsymbol{0},
\eqa
where $ \vH_T = \frac{1}{a} \boldsymbol{1}_a' \kp \vP_t$  and $\vH_{GT} = \vP_a \kp \vP_t.$
\een

We considered balanced as well as unbalanced designs for the $\vn =(n_1, n_2, n_3)$ subjects in group 1--3, respectively. 
The simulated numbers of subjects were $\vn^{(1)} =(30, 20, 10),\, \vn^{(2)} = (10, 20, 30)$ and $ \vn^{(3)} =(15, 15, 15)$. 
Furthermore, we simulated three different covariance structures $\vV_i$
\begin{itemize}
	\item[] Setting 1: $\vV_i = \vI_t$ for all $i=1, 2, 3$
	\item[] Setting 2: $\vV_i = diag(\sigma_1^2, \dots, \sigma_t^2)$ with $\sigma_s^2 = s$ for $t=4$ and $\sigma_s^2 = \sqrt{s}$ for $t=8$
	\item[] Setting 3: $\vV_i = \left( \rho_i^{|l-j|}\right)_{l,j \leq t}, (\rho_1,\rho_2,\rho_3)=(0.6, 0.5, 0.4)$ for $i=1, 2, 3$
%	\item[] Setting 4: $\vV_i = c_i \cdot \vI_t$ for $\vc = (1, \sqrt{2}, 2),~ i=1, 2, 3$.
\end{itemize}
In Setting 1 and 2 the covariance structures are the same for all groups, whereas in Setting~3 we have an autoregressive covariance structure 
with different parameters for the different groups. 
%The last setting describes a situation with increasing variances for $i=1,2,3,$ resulting in 
%negative pairing when combined with $\vn^{(1)}$ and in positive pairing when combined with $\vn^{(2)}$. 
Moreover, we simulated block effects 
with different variances $\sigma_i^2 \in \{0, 1, 2\}$. However, since the results were almost identical, we here only report the case $\sigma_i^2 = 0$.
All simulations were conducted with 10,000 simulation and 1,000 permutation runs. 

\subsection{Type-I error rates}
The resulting type-I error rates for the hypotheses of \emph{no time effect $T$} and 
\emph{no group $\times$ time interaction $GT$} are displayed in Tables~\ref{simu:resultsT} and \ref{simu:resultsGT}, respectively.

It is obvious that the tests based on the WTS considerably exceed the nominal level for small sample sizes. This behavior becomes worse with an 
increasing number of repeated measurements and when testing the interaction hypothesis. 
In some cases, the WTS reaches an empirical type-I error rate of almost 50\% when testing the $GT$-interaction. 
This means that its accuracy is no better than flipping a coin. 
The ATS, in contrast, keeps the pre-assigned level $\alpha$ pretty well for normally distributed observations, even for small sample sizes. With an increasing number of repeated measurements and/or non-normal data, 
however, the ATS leads to quite conservative decisions. Furthermore, the ATS leads to slightly conservative decisions when testing the interaction hypothesis, even with normally distributed data.
The WTPS is reasonably close to the pre-assigned level $\alpha$ in most situations, even under non-normality and for testing the interaction hypothesis. Despite the dependencies in longitudinal data, the permutation procedure greatly improves the behavior of the WTS in small sample settings. 
However, when testing the interaction hypothesis for $t=8$ repeated measurements the WTPS shows a 
more or less conservative behavior in Setting 3 combined with $\vn^{(2)}$, and a 
slightly liberal behavior for Setting 3 with $\vn^{(1)}$. 

%\newpage

%\thispagestyle{empty}
\begin{table}[H]
	\small
	\centering
	\caption{Results of the simulation studies for the hypothesis of no time effect.}
	\label{simu:resultsT}
	\begin{tabular}{c|c||ccc||ccc}
		\hline
		\multicolumn{8}{c}{normal distribution}\\ \hline \hline
		\multicolumn{2}{c|}{T} & \multicolumn{3}{c||}{$t=4$} & \multicolumn{3}{c}{$t=8$}\\ \hline
		Cov. Setting &	& ATS & WTS  & WTPS &  ATS &  WTS  & WTPS     \\ \hline
		\multirow{3}{*}{1} & $\vn^{(1)}$	&	0.046 & 0.085 & 0.050 & 0.040 & 0.177 & 0.050 \\ \cline{2-8}
		& $\vn^{(2)}$ &		0.046 & 0.086 & 0.048 & 0.040 & 0.177 & 0.052 \\  \cline{2-8}
		& $\vn^{(3)}$ & 		0.050 & 0.078 & 0.051 & 0.043 & 0.135 & 0.052 \\ \hline \hline
		\multirow{3}{*}{2} & $\vn^{(1)}$ &	0.051 & 0.085 & 0.050 & 0.042 & 0.177 & 0.051 \\ \cline{2-8}
		& $\vn^{(2)}$ & 	0.052 & 0.086 & 0.051 & 0.043 & 0.177 & 0.052 \\ \cline{2-8}
		& $\vn^{(3)}$ &	0.053 & 0.077 & 0.051 & 0.041 & 0.135 & 0.052 \\ \hline \hline
		\multirow{3}{*}{3} & $\vn^{(1)}$ &	0.046 & 0.092 & 0.052 & 0.044 & 0.198 & 0.062 \\ \cline{2-8}
		& $\vn^{(2)}$ &		0.051 & 0.080 & 0.045 & 0.048 & 0.155 & 0.042 \\ \cline{2-8}
		& $\vn^{(3)}$ &	0.051 & 0.078 & 0.053 & 0.048 & 0.136 & 0.054 \\ \hline \hline
%		\multirow{3}{*}{4} & $\vn^{(1)}$ &		0.047 & 0.104 & 0.065 & 0.036 & 0.254 & 0.099 \\ \cline{2-8}
%		& $\vn^{(2)}$ &	0.050 & 0.072 & 0.040 & 0.043 & 0.121 & 0.028 \\ \cline{2-8}
%		& $\vn^{(3)}$ &		0.050 & 0.084 & 0.056 & 0.040 & 0.148 & 0.063 \\ \hline
%		\hline
		\multicolumn{8}{c}{log-normal distribution}\\ \hline \hline
	%	\multicolumn{2}{c|}{T} & \multicolumn{3}{c||}{$t=4$} & \multicolumn{3}{c}{$t=8$}\\ \hline
		Cov. Setting &	& ATS & WTS  & WTPS &  ATS &  WTS  & WTPS     \\ \hline
		\multirow{3}{*}{1} & $\vn^{(1)}$	& 0.032 & 0.094 & 0.051 & 0.021 & 0.198 & 0.047 \\ \cline{2-8}
		& $\vn^{(2)}$ & 0.031 & 0.090 & 0.052 & 0.020 & 0.198 & 0.046 \\ \cline{2-8}
		& $\vn^{(3)}$ & 0.031 & 0.089 & 0.051 & 0.021 & 0.186 & 0.048 \\ \hline \hline
		\multirow{3}{*}{2} & $\vn^{(1)}$	& 0.040 & 0.110 & 0.067 & 0.022 & 0.207 & 0.053 \\ \cline{2-8}
		& $\vn^{(2)}$ & 0.040 & 0.107 & 0.067 & 0.022 & 0.203 & 0.051 \\ \cline{2-8}
		& $\vn^{(3)}$ & 0.042 & 0.107 & 0.070 & 0.024 & 0.197 & 0.057 \\ \hline \hline
		\multirow{3}{*}{3} & $\vn^{(1)}$	& 0.033 & 0.101 & 0.057 & 0.024 & 0.221 & 0.064 \\ \cline{2-8}
		& $\vn^{(2)}$ & 0.037 & 0.090 & 0.053 & 0.033 & 0.190 & 0.048 \\ \cline{2-8}
		& $\vn^{(3)}$ & 0.036 & 0.092 & 0.057 & 0.031 & 0.191 & 0.062 \\ \hline \hline
%		\multirow{3}{*}{4} & $\vn^{(1)}$	& 0.029 & 0.107 & 0.062 & 0.017 & 0.262 & 0.080 \\ \cline{2-8}
%		&  $\vn^{(2)}$  & 0.031 & 0.083 & 0.047 & 0.021 & 0.160 & 0.033 \\ \cline{2-8}
%		& $\vn^{(3)}$ & 0.032 & 0.091 & 0.053 & 0.020 & 0.197 & 0.058 \\ \hline
%		\hline
		\multicolumn{8}{c}{exponential distribution}\\ \hline \hline
	%	\multicolumn{2}{c|}{T} & \multicolumn{3}{c||}{$t=4$} & \multicolumn{3}{c}{$t=8$}\\ \hline
		Cov. Setting &	& ATS & WTS  & WTPS &  ATS &  WTS  & WTPS     \\ \hline
		\multirow{3}{*}{1} & $\vn^{(1)}$	& 0.045 & 0.090 & 0.048 & 0.034 & 0.194 & 0.051 \\ \cline{2-8}
		& $\vn^{(2)}$ &		0.046 & 0.096 & 0.053 & 0.032 & 0.191 & 0.048 \\ \cline{2-8}
		& $\vn^{(3)}$ &			0.046 & 0.086 & 0.054 & 0.034 & 0.151 & 0.050 \\ \hline \hline
		\multirow{3}{*}{2} & $\vn^{(1)}$	&	0.048 & 0.093 & 0.054 & 0.035 & 0.194 & 0.052 \\ \cline{2-8}
		& $\vn^{(2)}$ &		0.050 & 0.101 & 0.060 & 0.034 & 0.193 & 0.051 \\ \cline{2-8}
		& $\vn^{(3)}$ &			0.050 & 0.088 & 0.058 & 0.036 & 0.154 & 0.051 \\ \hline \hline
		\multirow{3}{*}{3} & $\vn^{(1)}$	&		0.049 & 0.098 & 0.055 & 0.042 & 0.218 & 0.066 \\ \cline{2-8}
		& $\vn^{(2)}$ &		0.050 & 0.090 & 0.049 & 0.046 & 0.173 & 0.045 \\ \cline{2-8}
		& $\vn^{(3)}$ &			0.050 & 0.087 & 0.055 & 0.042 & 0.153 & 0.056 \\ \hline \hline
%		\multirow{3}{*}{4} & $\vn^{(1)}$	&	0.044 & 0.111 & 0.064 & 0.028 & 0.274 & 0.096 \\ \cline{2-8}
%		& $\vn^{(2)}$ &		0.048 & 0.080 & 0.042 & 0.038 & 0.139 & 0.030 \\ \cline{2-8}
%		& $\vn^{(3)}$ &			0.046 & 0.088 & 0.056 & 0.035 & 0.168 & 0.058 \\ \hline
%		\hline
	\end{tabular}
\end{table}

\begin{table}[H]
	\small
	\centering
	\caption{Results of the simulation studies for the hypothesis of no group $\times$ time interaction.}
	\label{simu:resultsGT}
	\begin{tabular}{c|c|ccc||ccc}
		\hline
		\multicolumn{8}{c}{normal distribution}\\ \hline \hline
		\multicolumn{2}{c|}{GT} & \multicolumn{3}{c||}{$t=4$} & \multicolumn{3}{c}{$t=8$}\\ \hline
		Cov. Setting &	& ATS & WTS  & WTPS &  ATS &  WTS  & WTPS     \\ \hline
		\multirow{3}{*}{1} &  $\vn^{(1)}$	&	0.049 & 0.135 & 0.046 & 0.033 & 0.432 & 0.051 \\ \cline{2-8}
		&  $\vn^{(2)}$ &		0.053 & 0.142 & 0.052 & 0.034 & 0.433 & 0.050 \\ \cline{2-8}
		&  $\vn^{(3)}$ &	0.048 & 0.126 & 0.049 & 0.039 & 0.366 & 0.051 \\ \hline \hline
		\multirow{3}{*}{2} &  $\vn^{(1)}$	&		0.053 & 0.132 & 0.050 & 0.038 & 0.429 & 0.052 \\ \cline{2-8}
		&  $\vn^{(2)}$ &			0.053 & 0.141 & 0.054 & 0.038 & 0.431 & 0.050 \\ \cline{2-8}
		&  $\vn^{(3)}$ &		0.050 & 0.122 & 0.052 & 0.040 & 0.366 & 0.050 \\ \hline \hline
		\multirow{3}{*}{3} &  $\vn^{(1)}$	&		0.054 & 0.141 & 0.050 & 0.040 & 0.465 & 0.065 \\ \cline{2-8}
		&  $\vn^{(2)}$ &			0.053 & 0.135 & 0.045 & 0.049 & 0.393 & 0.037 \\ \cline{2-8}
		&  $\vn^{(3)}$ &		0.051 & 0.126 & 0.049 & 0.045 & 0.363 & 0.053 \\ \hline \hline
%		\multirow{3}{*}{4} &  $\vn^{(1)}$	&			0.053 & 0.154 & 0.062 & 0.034 & 0.539 & 0.121 \\ \cline{2-8}
%		&  $\vn^{(2)}$ &			0.051 & 0.122 & 0.041 & 0.035 & 0.317 & 0.018 \\ \cline{2-8}
%		&  $\vn^{(3)}$ &		0.050 & 0.131 & 0.055 & 0.037 & 0.382 & 0.061 \\ \hline
%		\hline
		\multicolumn{8}{c}{log-normal distribution}\\ \hline \hline
%		\multicolumn{2}{c|}{GT} & \multicolumn{3}{c||}{$t=4$} & \multicolumn{3}{c}{$t=8$}\\ \hline
		Cov. Setting &	& ATS & WTS  & WTPS &  ATS &  WTS  & WTPS     \\ \hline
		\multirow{3}{*}{1} &  $\vn^{(1)}$	& 0.024 & 0.121 & 0.047 & 0.012 & 0.426 & 0.053 \\ \cline{2-8}
		&  $\vn^{(2)}$ &		0.022 & 0.128 & 0.053 & 0.013 & 0.431 & 0.051 \\ \cline{2-8}
		&  $\vn^{(3)}$ &		0.024 & 0.118 & 0.048 & 0.012 & 0.406 & 0.051 \\ \hline \hline
		\multirow{3}{*}{2} &  $\vn^{(1)}$	&	0.025 & 0.129 & 0.051 & 0.014 & 0.427 & 0.054 \\ \cline{2-8}
		&  $\vn^{(2)}$ &		0.026 & 0.130 & 0.054 & 0.013 & 0.432 & 0.052 \\ \cline{2-8}
		&  $\vn^{(3)}$ &		0.023 & 0.120 & 0.050 & 0.013 & 0.403 & 0.052 \\ \hline \hline
		\multirow{3}{*}{3} &  $\vn^{(1)}$	&		0.029 & 0.133 & 0.050 & 0.020 & 0.457 & 0.062 \\ \cline{2-8}
		&  $\vn^{(2)}$ &		0.028 & 0.121 & 0.045 & 0.024 & 0.399 & 0.036 \\ \cline{2-8}
		&  $\vn^{(3)}$ &		0.028 & 0.122 & 0.049 & 0.020 & 0.408 & 0.053 \\ \hline \hline
%		\multirow{3}{*}{4} &  $\vn^{(1)}$	&		0.025 & 0.147 & 0.065 & 0.012 & 0.544 & 0.121 \\ \cline{2-8}
%		&  $\vn^{(2)}$ &		0.024 & 0.108 & 0.044 & 0.015 & 0.335 & 0.026 \\ \cline{2-8}
%		&  $\vn^{(3)}$ &		0.025 & 0.128 & 0.054 & 0.012 & 0.427 & 0.067 \\ \hline
%		\hline
		\multicolumn{8}{c}{exponential distribution}\\ \hline \hline
%		\multicolumn{2}{c|}{GT} & \multicolumn{3}{c||}{$t=4$} & \multicolumn{3}{c}{$t=8$}\\ \hline
		Cov. Setting &	& ATS & WTS  & WTPS &  ATS &  WTS  & WTPS     \\ \hline
		\multirow{3}{*}{1} &  $\vn^{(1)}$	& 0.043 & 0.146 & 0.054 & 0.024 & 0.442 & 0.054 \\ \cline{2-8}
		&  $\vn^{(2)}$ &		0.041 & 0.148 & 0.054 & 0.024 & 0.443 & 0.050 \\ \cline{2-8}
		&  $\vn^{(3)}$ &		0.036 & 0.122 & 0.047 & 0.028 & 0.397 & 0.054 \\ \hline \hline
		\multirow{3}{*}{2} &  $\vn^{(1)}$	&	0.048 & 0.151 & 0.059 & 0.027 & 0.444 & 0.057 \\ \cline{2-8}
		&  $\vn^{(2)}$ &		0.042 & 0.153 & 0.059 & 0.025 & 0.448 & 0.052 \\ \cline{2-8}
		&  $\vn^{(3)}$ &		0.034 & 0.121 & 0.048 & 0.029 & 0.397 & 0.055 \\ \hline \hline
		\multirow{3}{*}{3} &  $\vn^{(1)}$	&		0.047 & 0.155 & 0.061 & 0.032 & 0.473 & 0.068 \\ \cline{2-8}
		&  $\vn^{(2)}$ &		0.043 & 0.140 & 0.049 & 0.042 & 0.406 & 0.037 \\ \cline{2-8}
		&  $\vn^{(3)}$ &		0.037 & 0.122 & 0.047 & 0.041 & 0.402 & 0.058 \\ \hline \hline
%		\multirow{3}{*}{4} &  $\vn^{(1)}$	&		0.043 & 0.177 & 0.076 & 0.023 & 0.559 & 0.134 \\ \cline{2-8}
%		&  $\vn^{(2)}$  &		0.041 & 0.123 & 0.043 & 0.028 & 0.334 & 0.022 \\ \cline{2-8}
%		&  $\vn^{(3)}$  &		0.036 & 0.128 & 0.053 & 0.030 & 0.418 & 0.068 \\ \hline
%		\hline
	\end{tabular}
\end{table}

\newpage
The simulations show a clear advantage of the permutation procedure as compared to the $\chi^2$- approximation of the Wald-type statistic. The WTPS controlled the 5\% level in most situations, even under non-normality, i.e., in situations where the ATS 
may lead to quite conservative decisions.

\subsection{Additional simulation results}
We note that additional simulations for the type-I error can be found in the supplementary material to this paper. 
There we have compared the above methods with other resampling schemes such as the bootstrap procedures described in \cite{Kon:2015}. 
Of all procedures analyzed in the simulations, the permutation procedure produced the best results.

\subsubsection{Quality of the approximation}

\begin{figure}[h]
		\centering
		\includegraphics[width=0.7\textwidth]{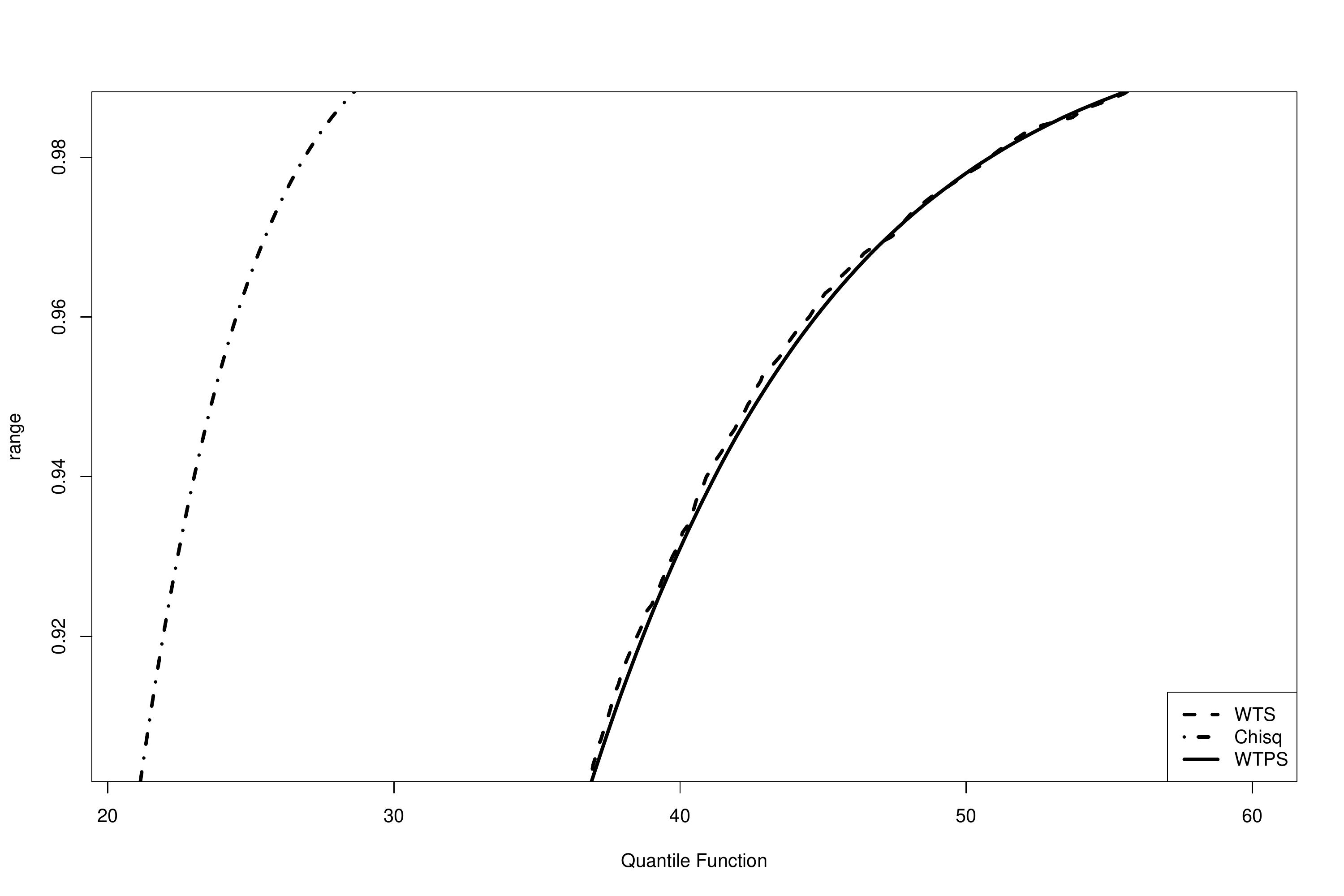}
		\caption{\it Quantile funtions of the WTS, WTPS and the corresponding $\chi^2$-distribution in the balanced simulation setting with log-normally distributed data, $t=8$, covariance matrix setting 2 and under the null hypothesis of no interaction.}
		\label{Fig.QF_logn}
\end{figure}

In the following, we denote by $F_N$ the distribution function of $Q_N$ under $H_0$, by $F$ the distribution function of the limiting $\chi^2_f$-distribution under $H_0$ and by $F_N^{\pi}$ the distribution function of the WTPS under $H_0$. We can now define
$
KQS = \sup_{0.9 \leq t \leq 0.99} | F_N^{-1}(t) - F^{-1}(t)|
$
as well as
$
KQS^\pi = \sup_{0.9 \leq t \leq 0.99} | F_N^{-1}(t) - (F_N^\pi)^{-1}(t)|
$
in order to compare the distance between the quantile function $F_N^{-1}$ and the limiting quantile function $F^{-1}$ ($KQS$) with the distance between $F_N^{-1}$ and $(F_N^\pi)^{-1}$, the quantile functions of the test statistic and its permuted version ($KQS^\pi$), respectively. We have calculated these distances for all simulation settings described above. Detailed results can be found in Section \ref{supp:AdditionalResults} of the supplementary material. It turned out that $KQS^\pi$ is always smaller than $KQS$, i.e., the approximation provided by the permutation procedure is considerably better than the asymptotic $\chi^2$ approximation for all simulation settings considered. In our simulations, $KQS$ ranged from 1.991 to 48.11 with a median distance of 9.179, whereas $KQS^\pi$ ranged from  0.1049 to 7.618 with a median value of 0.8948.
Figure \ref{Fig.QF_logn} exemplarily shows the plots of the corresponding quantile functions for one of the simulation scenarios.

\subsubsection{Large sample behavior}

\begin{figure}[H]
	\centering
	\includegraphics[width=0.8\textwidth]{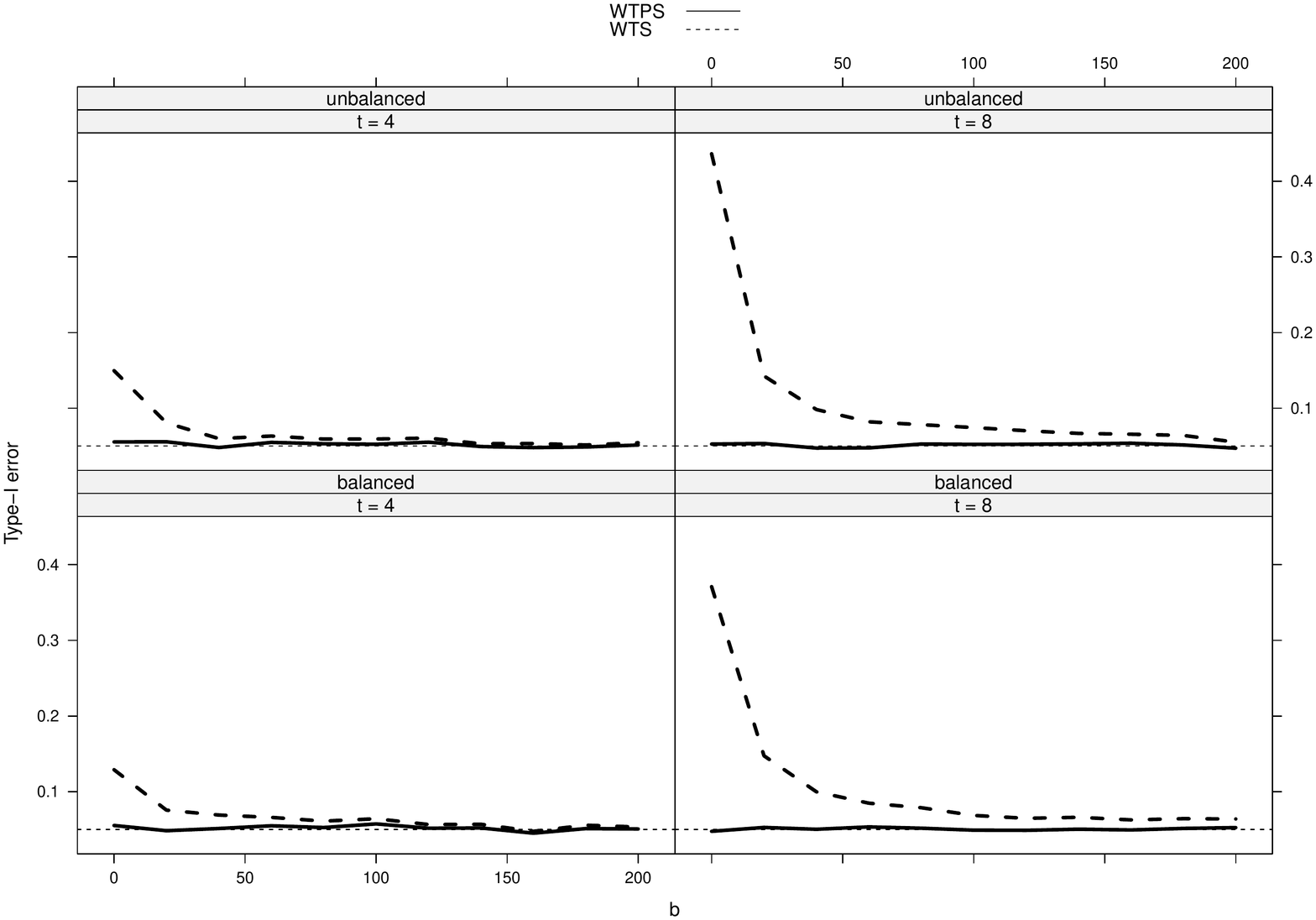}
	\caption{\it Type-I error rates under the interaction hypothesis for the WTS and the WTPS, where sample size was increased by adding $b\veins_3, b=0, 20, \dots, 200$ to the sample size vectors in a balanced (lower panel) and unbalanced (upper panel) design with $t=4$ (left panel) and $t=8$ (right panel) time points under covariance setting 2, i.e., $\vV_i = diag(\sigma_1^2, \dots, \sigma_t^2)$ with $\sigma_s^2 = s$ for $t=4$ and $\sigma_s^2 = \sqrt{s}$ for $t=8$.}
	\label{Fig.large_vari}
\end{figure}

In this section, we analyze the large sample behavior of the WTS and WTPS. We considered only normally distributed random variables with covariance structure Setting 2 for an unbalanced ($\vn^{(1)}=(30, 20, 10)$) as well as a balanced ($\vn^{(3)}=(15, 15, 15)$) design with $t=4, 8$ time points. 
The sample size was increased by adding $b \veins_3$ to the above sample size vectors for $b=0, 20, \dots, 200$. The results for the type-I error under the null hypothesis of no interaction and covariance setting 2 are presented in Figure \ref{Fig.large_vari}.
The behavior of the WTS improves with growing sample size but even for 115 individuals in all groups, the WTS still exceeds the nominal level. The WTPS, in contrast, is rather close to the pre-assigned level even for small sample sizes.
%Further results obtained for covariance setting 4 can be found in Section \ref{supp:AdditionalResults} of the supplementary material.

\subsubsection{Power}

The power simulations are explained in detail in Section \ref{supp:Power} of the supplementary material to this paper. Since the WTS turned out to test on different $\alpha-$levels (see the simulation results under the null hypothesis), we have excluded it from the analyses. We additionally considered the approximation described by \citet{Lecoutre1991} as well as Hotelling's $T^2$ \citep{Hotelling:1931}.
It turns out that the ATS has the highest power for normally distributed data, performing slightly better than the WTPS. 
For log-normally distributed data, the WTPS has larger power than the other methods and it is the only method controlling the type-I error correctly.

\section{Application: Analysis of the Data Example}\label{app}

Finally, we analyze the data example on oxygen consumption of leukocytes in the presence and absence of inactivated staphylococci. In this setting we wish to analyze the effect of the whole-plot factor 'treatment' (factor A, Placebo/Verum, $a=2$) as well as the sub-plot factors 'staphylococci' (factor B, with/without, $b=2$) and 'time' (factor T, 6/12/18 min, $t=t_i=3, i=1, \dots, ab$). We are also interested in interactions between the different factors. 
The mean values of the data are given in Table \ref{bsp.leuko2.tab} in Section~\ref{int}. 
\iffalse	\begin{table}[h] 
		\centering
		\caption{\it Mean oxygen consumption of leukocytes in the presence and absence of inactivated staphylococci.}
		\label{bsp.leuko.tab} 
		{\small
			\begin{tabular}{|c||rrr|rrr|} \hline
				\multicolumn{7}{|c|}{Mean $O_2$-consumption [$\mu \ell$]} \\ \hline
				& \multicolumn{6}{c|}{staphylococci} \\
				& \multicolumn{3}{c|}{with} & \multicolumn{3}{c|}{without} \\ \hline
		Treatment		& \multicolumn{3}{c|}{Time [Min]} & \multicolumn{3}{c|}{Time [Min]} \\
				& 6 & 12 & 18 & 6 & 12 & 18  \\ \hline \hline
				Placebo & 1.618 & 2.434 & 3.527 & 1.322 & 2.430& 3.425 \\[-1mm] \hline
				Verum & 1.656 & 2.799 & 4.029 & 1.394 & 2.57 & 3.677 \\[-1mm]
				\hline
				\hline
			\end{tabular}
		}
	\end{table}
 \fi
 %Figure \ref{Fig.boxplot} shows the boxplots of the data in the two treatment groups Placebo and Verum, each in the presence and absence of staphylococci. The dots denote the mean values. The plots indicate an effect of all three factors, with the time effect being most pronounced.

In the analysis we compared the three tests discussed above: The ATS in \eqref{ATSstatistic} is compared to the corresponding 
$F(\hat{\nu},\infty)$-quantile, 
the WTS in \eqref{WTS} to the asymptotic $\chi^2_f$-quantile as well as the quantile obtained by the permutation procedure (WTPS). 
The seven different null hypotheses of interest about main and interaction effects can be tested by 
choosing the related hypotheses matrices. Here, we have chosen 
$\vH_A=\vP_a \otimes \frac{1}{b} \boldsymbol{1}_b' \otimes \frac{1}{t} \boldsymbol{1}_t'$, $\vH_B=\frac{1}{a} \boldsymbol{1}_a' \otimes \vP_b \otimes \frac{1}{t} \boldsymbol{1}_t'$ and $\vH_T=\frac{1}{a} \boldsymbol{1}_a' \otimes \frac{1}{b} \boldsymbol{1}_b' \otimes \vP_t$ 
for testing the main effect of the three factors $A,B,$ and $T$. 
For the interaction terms we used the matrices $\vH_{AT}=\vP_a \otimes \frac{1}{b} \boldsymbol{1}_b' \otimes \vP_t$, $\vH_{AB}=\vP_a \otimes \vP_b \otimes \frac{1}{t} \boldsymbol{1}_t'$ and $\vH_{BT}=\frac{1}{a} \boldsymbol{1}_a' \otimes \vP_b \otimes \vP_t,$ and $\vH_{ABT} = \vP_a \otimes \vP_b \otimes \vP_t$. 
The resulting p-values of the analysis are presented in Table \ref{table:data}. 
\iffalse
 \begin{figure}[h]
 	\centering
 	\includegraphics[width=0.5\textwidth]{Rplot02.pdf}
 	\caption{\it Boxplots of the $O_2$ consumption data for the different treatment groups (upper plots: Placebo, lower plots: Verum) with and without staphylococci, respectively. The dots denote the mean values in each group for the three different time points.}
 	 	\label{Fig.boxplot}
 \end{figure}
  \fi

\begin{table}[h] 
	\centering
		\caption{\it Results of the analysis of the $O_2$ consumption data.}
		\label{table:data}
		\vspace{0.3cm}
	\begin{tabular}{r|rrr}
		\hline
		& ATS & WTS &  WTPS   \\ 
		\hline
		A  &  0.001 & 0.001 &  $<$0.001  \\ 
		B &  $<$0.001 &  $<$0.001 & $<$0.001  \\ 
		T & $<$0.001 &  $<$0.001 &  $<$0.001  \\ 
		AB & 0.110 & 0.110 &  0.118  \\ 
		AT &  0.009  & $<$0.001 &  0.001 \\ 
		BT &  0.094 & 0.115 &  0.141  \\ 
		ABT  &  0.117 & 0.116 & 0.156  \\ 
		\hline
	\end{tabular}
\end{table}

For this example all tests under considerations lead to similar conclusions: 
Each factor (treatment, staphylococci and time) has a significant influence on the $O_2$ consumption of the leukocytes. 
Moreover, there is a significant interaction between treatment and time.

% dis.tex

\section{Conclusions and Discussion} \label{dis}

In this paper, we have generalized the permutation idea of \cite{PBK} for independent univariate factorial designs to the case of repeated measures 
allowing for a factorial structure. 
Here, the suggested permutation test is asymptotically valid and does not require the 
assumptions of multivariate normality, equal covariance matrices or balanced designs. 
It is based on the well-known Wald-type statistic (WTS) which possesses the 
beneficial property of an asymptotic pivot while being applicable for general repeated measures designs. 
Since it is well known for being very liberal for small and moderate sample sizes, we have considerably improved its 
small sample behavior under the null hypothesis by a studentized permutation technique. 
For univariate and independent observations the idea of this technique dates back to \cite{Neuhaus:1993} and 
\cite{janssen:1997} and has recently been considered for more complex designs in independent observations by \cite{Chung:2013} and \cite{PBK}.

In addition, we have rigorously proven in Theorem~\ref{theo} 
that the permutation distribution of the WTS always approximates the null distribution of the WTS and can thus be applied for calculating data-dependent critical values. In particular, the result implies that the corresponding Wald-type permutation test is asymptotically exact under the null hypothesis 
and consistent for fixed alternatives while providing the same local power as the WTS under contiguous alternatives.

Moreover, our simulation study indicated that the permutation procedure showed a very accurate performance in all designs under consideration with moderate repeated measures (t=4) and homoscedastic or slightly heteroscedastic covariances. 
Only in the case of a larger number of repeated measurements (t=8) the 
WTPS showed a more or less liberal (conservative) behavior when testing the interaction hypothesis in an unbalanced design. 
However, all other competing procedures considered in the paper and the supplementary material did not perform 
better in these situations.

Roughly speaking, the good performance of the WTPS for finite samples may be explained by a better approximation of the underlying distribution of the WTS by the permutation distribution as compared to the $\chi^2$-distribution. This could be seen clearly in the distances between the quantile functions.

%{\color{blue} Noch irgendwo einfügen(?): As pointed out by \cite{Kon:2015} ``The investigation of permutation methods in general
%multivariate factorial designs, however, is challenging and will be part of future research.''}

\section*{Acknowledgement} 
The authors would like to thank Frank Konietschke for helpful discussions.\\
This work was supported by the German Research Foundation project DFG-PA 2409/3-1.

% app.tex

%\textit{Proof of Remark \ref{rem:2}:}
%From Theorem \ref{theo} it follows, that $c_{1-\alpha}^*$ converges to $\chi^2_{f, 1-\alpha}$ in probability and therefore $\varphi_{WTPS} - \varphi_{WTS} \stackrel{P}{\to} 0$ as $N \to \infty$.
%
%...
%
%\qed

\newpage
\begin{center}
\title{\Large \bf Supplement To\\
	Permuting Longitudinal Data Despite All The Dependencies\\
}
\author{Sarah Friedrich$^{*}$,  Edgar Brunner$^{**}$ and  Markus Pauly$^{*}$ \\[1ex] 
	%{\small University of Gï¿½ttingen, Germany}
}
\begin{abstract}
	In this supplementary material to the authors' paper 'Permuting longitudinal data despite all the dependencies' 
	we present all technical details together with additional simulation results and consider further resampling techniques, 
	namely a nonparametric and a parametric bootstrap approach. 
	Furthermore, we analyze the power of the WTPS and compare it to the power of the ATS. 
	Finally, we present more details about the the data example on $O_2$-consumption of leukocytes 
	and also present an analysis, where we applied all considered approaches.
\end{abstract}
\end{center}

\vfill
\vfill

\noindent${}^{*}$ {University of Ulm, Institute of Statistics, Germany\\
	\mbox{ }\hspace{1 ex}email: markus.pauly@uni-ulm.de}

\noindent${}^{**}$ {University Medical Center G\"ottingen, Institute of Medical Statistics, Germany}

\newpage

\setcounter{section}{7}
\setcounter{table}{5}

\section{Proofs}

\textit{Proof of Theorem \ref{theo:ATS}:}
First note that $\vT=\vT'$ as well as $\vT^2=\vT$. Let $\vT \vmu = \frac{\vT \vnu}{\sqrt{N}}$ for $\vnu \in \renu^T$, i.e., for $\vnu = \vnull$ we are working under $H_0$. It holds that $\sqrt{N}(\volY_{\cdot} - \vmu)$ has, asymptotically, 
a multivariate normal distribution with mean $\vnu$ and covariance matrix $\vSigma$. Thus, it follows that 
\[
N \volY_{\cdot}'\vT \volY_{\cdot} \to \vZ' \vT \vZ,
\]
with $\vZ \sim N(\vnu, \vSigma)$. If additionally $\vSigma >0$, we may write $\vZ = \vSigma^{1/2} \tilde{\vZ}$ where $\tilde{\vZ} \sim N(\vSigma^{-1/2}\vnu, \vI)$ and thus $\vZ' \vT \vZ = \sumi a \sums {t_i} \lambda_{is} X_{is}$ where $\lambda_{is}$ are the eigenvalues of $\vT \vSigma$ and $X_{is} \sim \chi^2_1(\delta)$ for $\delta= \vnu' \vSigma^{-1}\vnu$.
\qed

\textit{Proof of Theorem \ref{WTS_asy}:}
The null distribution of the WTS follows analogous to the proof of Theorem~2.1 in \cite{Kon:2015}. Obviously, $\varphi_{WTS}$ is an asymptotic level $\alpha$ test and consistent for fixed alternatives $\vH \vmu \neq \vnull$.\\
Under $H_1: \vH \vmu = \frac{1}{\sqrt{N}}\vnu$, it holds that $\sqrt{N} \vH \volY_{\cdot}$ has, asymptotically, an $N(\vH \vnu, \vH \vSigma \vH')$ distribution. Thus, the WTS has asymptotically a non-central $\chi^2_f(\tilde{\delta})$ distribution with $f=rank(\vH)$ degrees of freedom and non-centrality parameter $\tilde{\delta} = (\vH \vnu)' (\vH \vSigma \vH')^+ \vH \vnu$. 
\qed

We will now proof Theorem \ref{theo}.
For notational convenience, we introduce 
\bqa
\vZ = (Z_{N, 1}, \dots, Z_{N, \tilde{N}}) = (Y_{111}, Y_{121}, \dots, Y_{1n_11}, Y_{112}, \dots, Y_{an_at_a})
\eqa
for the pooled sample. Since $\vH \boldsymbol{1} = \boldsymbol{0}$ we can rewrite the permuted test statistic as
\bqa
Q_N^{\pi} =\sqrt{ N} (\volY_{\cdot}^{\pi}-\volY_{\ldots})'\vH'(\vH \vwhSigma^{\pi} \vH')^+\sqrt{N}\vH (\volY_{\cdot}^{\pi}-\volY_{\ldots}),
\eqa
where $\volY_{\ldots}=\olY_{\ldots} \cdot \boldsymbol{1}_{T} $ and $\olZ_{\tilde{N}} = \olY_{\ldots} = \frac{1}{\tilde{N}} \sumi {\tilde{N}} {Z_{N, i}}$.
Based on this representation, we can split the proof of Theorem \ref{theo} in two results. 
There, we first show that the conditional distribution of $\sqrt{N} (\volY_{\cdot}^{\pi}-\volY_{\ldots})$ given the data is 
asymptotically multivariate normal. However, it turns out that the resulting covariance matrix is different from $\vSigma$. 
Our approach corrects for the 'wrong' covariance structure by studentizing with $\vwhSigma^{\pi}$, 
which is shown in a second step. Altogether, this proves the consistency of the WTPS as stated in Theorem \ref{theo} as well as the properties 
of the corresponding test mentioned in Remarks~\ref{rem:1} and \ref{rem:2}.

Note that there exist finite limits $b_i = \lim_{\min(n_i)\to \infty} \frac{\tilde{N}}{n_i} \in (1, \infty), i=1, \dots, a$ because of (1) and $0 < \max_{i=1, \dots, a}(t_i) < \infty$.

\begin{lemma} \label{1}
	Under the assumptions of Theorem \ref{theo}, the conditional permutation distribution of
	\bqa
	\sqrt{N} (\volY_{\cdot}^{\pi}-\volY_{\ldots})
	\eqa
	given the observed data $\vY$ weakly converges to a multivariate normal --- $N(\vnull, \sigma^2\vGamma)$ --- distribution in probability, where
	\bqan \label{sigma}
	\sigma^2 = \sumi a \frac{1}{b_i} \sums {t_i} (\sigma^2_{i s}+ \mu^2_{i s})- \left(\sumi a \frac{1}{b_i} \sums {t_i} \mu_{i s}\right)^2
	\eqan
	with $\sigma_{is}^2 = \var(Y_{iks})$ and
	\bqan
	\vGamma = \bks_{i=1}^a \kappa_i ^{-1} \vI_{t_i} - \vJ_{T} = 
	\diag(\kappa_1^{-1}\vI_{t_1}, \dots, \kappa_a^{-1}\vI_{t_a}) - \vJ_{T}.
	\eqan
\end{lemma}

\proofit First note that the classical Cram\'{e}r-Wold device cannot be applied directly in this context due to the occurrence of uncountably many exceptional sets. Therefore we will apply a modified Cram\'{e}r-Wold device, see e.g.~the proof of Theorem 4.1 in \cite{Pauly:2011a}. Let $D$ be a dense and countable subset of $\renu^{T}$. Then for every fixed $\vlambda = (\lambda_1, \dots, \lambda_{T}) \in D$ and $M_0 = 0, M_1 = n_1, \dots, M_{t_1} = t_1 n_1, M_{t_1+1} = t_1n_1 +n_2, \dots, M_{T} =\tilde{N}$, we have
\bqa 
\sqrt{N}\vlambda' \volY_{\cdot} &=& \sumi a \sum_{k=M_{i-1}+1}^{M_i} \frac{\sqrt{N}}{n_i} \lambda_i Z_{N, k}\\
&=& \sqrt{N} \sums {\tilde{N}}c_{Ns} \frac{Z_{N,s}}{\sqrt{\tilde{N}}},
\eqa
where $c_{Ns} = \sqrt{\tilde{N}} \sumi {T} \1 \{M_{i-1}+1 \leq s \leq M_i \} \frac{\lambda_i}{n_i}$. This implies
\bqan \label{perm:diff}
\sqrt{N} \vlam' (\volY_{\cdot}^{\pi}- \volY_{\ldots}) &=& \sqrt{N} \sums {\tilde{N}} c_{Ns} \frac{(Z_{N, \pi(s)}- \olZ_{\tilde{N}})}{\sqrt{\tilde{N}}}\\
&\stackrel{d}{=}& \sqrt{N} \sums {\tilde{N}} c_{N\pi(s)}\frac{Z_{N, s}- \olZ_{\tilde{N}}}{\sqrt{\tilde{N}}}, \notag
\eqan
since $\pi$ is uniformly distributed on the set of all permutations of the numbers $\{1, \dots, \tilde{N}\}$. Let $b_i = \lim_{\min(n_i) \to \infty}\frac{\tilde{N}}{n_i}$ with $b_i < \infty$ because of (1) and $\max(t_i) < \infty$.

We now apply Theorem 4.1 in \cite{Pauly:2011a} to prove the conditional convergence in distribution. Therefore, we have to prove the following conditions:

\bqan 
\frac{1}{\sqrt{\tilde{N}}} \max _{1 \leq i \leq \tilde{N}} | Z_{N, i}-\olZ_{\tilde{N}}| &\stackrel{P}{\to}& 0 \label{cond:1}\\
\frac{1}{\tilde{N}} \sumi {\tilde{N}} (Z_{N,i}-\olZ_{\tilde{N}})^2 &\stackrel{P}{\to}& \sigma^2 \label{cond:2}\\
\max_{1 \leq s \leq \tilde{N}} |c_{N,s} - \olc_{\cdot}| &\stackrel{P} {\to}& 0 \label{cond:3}\\
\sums {\tilde{N}} (c_{N, s} - \olc_{\cdot})^2 &\stackrel{P}{\to}& \sigma^2_{\lambda} = \sumi T \lambda_i^2 b_i - \left(\sumi T \lambda_i\right)^2 \label{cond:4}\\
\sqrt{\tilde{N}} (c_{N,\pi(1)}-\olc_{\cdot}) &\stackrel{d}{\to}& W \text{ with } E(W)=0 \text{ and } \Var(W)=\sigma_{\lam}^2 \label{cond:5} 
\eqan
Condition \eqref{cond:1} as well as \eqref{cond:3} -- \eqref{cond:5} follow analogous to \cite{PBK}:
Since the random variables within each of the $a$ groups are i.i.d.~with finite variance, they fulfill \eqref{cond:1}.
The convergence in \eqref{cond:3} is obvious and since $\sqrt{\tilde{N}} \cdot \olc_{\cdot} = \sumi T \lam_i $ we have
\bqa
\sums {\tilde{N}} (c_{N, s}-\olc_{\cdot})^2 &=& \sums {\tilde{N}} c_{N, s}^2 - (\sqrt{\tilde{N}} \olc_{\cdot})^2= \sumi T\frac{\tilde{N}}{n_i} \lam_i^2 -\left(\sumi T\lam_i\right)^2\\
&\stackrel{P}{\to} & \sumi T \lam_i^2 b_i- \left(\sumi T \lam_i\right)^2  = \sigma_{\lam}^2.
\eqa
Moreover, \eqref{cond:5} holds due to
\bqa
P\left(\sqrt{\tilde{N}} c_{N,\pi(1)}=\frac{\tilde{N} \lam_i}{n_i}\right) = \frac{n_i}{\tilde{N}} \to \frac{1}{b_i} 
\eqa
for $1 \leq i \leq a$, i.e., for a random variable $\wtW$ with $P(\wtW = b_i \lam_i)=\frac{1}{b_i}$, $1 \leq i \leq a$, we have
\bqa
\sqrt{\tilde{N}} (c_{N, \pi(1)}-\olc_{\cdot}) \stackrel{d}{\to} \wtW - \sumi T \lam_i = W,
\eqa
where $W$ fulfills $E(W)=0$ and $\Var(W)=\sigma_{\lam}^2$.
It remains to prove \eqref{cond:2}:
\bqa
\frac{1}{\tilde{N}} \sumi {\tilde{N}} (Z_{N,i}-\olZ_{\tilde{N}})^2 = \frac{1}{\tilde{N}} \sumi {\tilde{N}} Z_{N,i}^2 -\olY_{\ldots}^2.
\eqa
Consider
\bqa
E \left(\frac{1}{\tilde{N}} \sumi {\tilde{N}} Z_{N,i}^2 \right) &=&  \frac{1}{\tilde{N}}\sumi a  \sums {t_i} \sumk {n_i} E \left(Y_{iks}^2 \right)\\
&=& \frac{1}{\tilde{N}}\sumi a  \sums {t_i} \sumk {n_i} (\sigma_{is}^2 + \mu_{is}^2)\\
&=& \sumi a \frac{n_i}{\tilde{N}} \sums {t_i} (\sigma_{i s}^2 + \mu_{i s}^2)\\
&\to& \sumi a \frac{1}{b_i} \sums{t_i} (\sigma_{is}^2 +  \mu_{i s}^2).
\eqa
Furthermore:
\bqan \label{EwertYbar}
E(\olY_{\ldots}^2) &=& E\left(\left(\frac{1}{\tilde{N}} \sumi {\tilde{N}} Z_{N,i}\right)^2\right)  \notag \\
&=& \underbrace{\var\left(\frac{1}{\tilde{N}} \sumi a \sums {t_i} \sumk {n_i} Y_{iks}\right)}_{{\mathcal O}(\frac{1}{N})} +  \left(E\left(\frac{1}{\tilde{N}} \sumi a \sums {t_i} \sumk {n_i} Y_{iks}\right)\right)^2\\ 
&\to& \left(\sumi a \frac{1}{b_i} \sums {t_i} \mu_{is}\right)^2 \notag
\eqan
Since
\bqa
\Var\left(\frac{1}{\tilde{N}} \sumi {\tilde{N}} (Z_{N,i} - \olZ_{\tilde{N}})^2\right) &=& \Var \left(\sumi a \frac{1}{\tilde{N}}\sums {t_i} \sumk {n_i} Y_{iks}^2 - \olY_{\ldots}^2 \right)\\
&=& \sumi a \frac{1}{(\tilde{N})^2} \sumk {n_i} \Var\left( \sums {t_i} (Y_{iks}^2 - \frac{1}{\tilde{N}}\olY_{\ldots}^2)\right)\\
&=& {\mathcal O}\left(\frac{1}{N}\right)
\eqa
because of independence and condition (2), the desired conclusion follows with Tschebyscheff's inequality.

Altogether, this implies by Theorem 4.1 in \cite{Pauly:2011a} convergence in distribution given the data $\vY$
\bqan \label{lambdaconvprob}
\sqrt{N} \vlam' (\volY_{\cdot}^{\pi}- \volY_{\ldots}) \stackrel{d}{\to} N(0, \sigma^2\sigma_{\lambda}^2)
\eqan
in probability. This convergence holds for every fixed $\vlambda \in D$. Applying the subsequential principle for convergence in probability we can find a common subsequence such that \eqref{lambdaconvprob} holds almost surely for all $\vlambda \in D$ along this subsequence. Now continuity of the characteristic function of the limit and tightness of the conditional distribution of $\sqrt{N} (\volY_{\cdot}^{\pi}- \volY_{\ldots})$ given $\vY$ show that \eqref{lambdaconvprob} holds almost surely for all $\vlambda \in \renu^T$ along this subsequence. Thus, an application of the classical Cram\'{e}r-Wold device together with another application of the subsequence principle imply the result.
\qed

Now we will study the convergence of $\vwhSigma^\pi = N \vwhV^\pi = \bks_{i=1}^a \frac{N}{n_i} \vwhV_i^\pi$. 
\begin{lemma} \label{2}
	Under the assumptions of Theorem \ref{theo} we have convergence in probability
	\bqa
	\vwhSigma^\pi \stackrel{P}{\to} \sigma^2 \diag(\kappa_1^{-1}\vI_{t_i}, \dots, \kappa_a^{-1}\vI_{t_a})
	\eqa
	as $N \to \infty$.
\end{lemma}

\proofit
It suffices to show that $(\vwhV_i^\pi)_{r, s} \stackrel{P}{\to} \sigma^2\1\{r=s\}$ in probability for all $1 \leq r, s \leq t_i$.
Therefore consider
\bqa
\frac{n_i-1}{n_i}(\vwhV_i^\pi)_{r, s} = \underbrace{\frac{1}{n_i} \sumk {n_i} Y_{ikr}^\pi Y_{iks}^\pi}_{A} -\underbrace{ \olY_{i \cdot r}^\pi \olY_{i \cdot s}^\pi}_{B}
\eqa
First, consider $B$. It holds:
\bqa
E(\olY_{i \cdot r}^\pi|\vY) \stackrel{P}{\to} \sumi a \frac{1}{b_i} \sums {t_i} \mu_{is}
\eqa
for all $r$ and all $i$ analogous to \eqref{EwertYbar}.
Furthermore, setting $d_{N,s}^{(i)} := \1\{(r-1)n_i+1 \leq s \leq rn_i\}/n_i$ for $1 \leq s \leq \tilde{N}$ and using Theorem 3 from \cite{Hajek} we get convergence in probability of the corresponding conditional variance
\bqa
\Var \left(\olY_{i \cdot r}^\pi| \vY\right) &=& \Var \left(\sums {\tilde{N}} d_{N, s}^{(i)} Z_{N, \pi(s)} | \vZ \right)\\
&=& \sums {\tilde{N}} \left(d_{N, s}^{(i)}- \old_{N, \cdot}^{(i)}\right)^2 \frac{1}{\tilde{N}-1}\sums {\tilde{N}} (Z_{N, s} - \olZ_{\tilde{N}})^2 \stackrel{P}{\to} 0,
\eqa
since $\sums {\tilde{N}} \left(d_{N, s}^{(i)}- \old_{N, \cdot}^{(i)}\right)^2 \to 0$ as $N \to \infty$ and $\frac{1}{\tilde{N}-1}\sums {\tilde{N}} (Z_{N, s} - \olZ_{\tilde{N}})^2 = \mathcal{O}_P(1)$. Altogether this implies convergence in probability by the continuous mapping theorem
\bqa
B \stackrel{P}{\to} \left(\sumi {a} \frac{1}{b_i} \sums {t_i} \mu_{i s}\right)^2.
\eqa

For part $A$ we distinguish two cases: First, assume $r=s$. We have
\bqa
A = \frac{1}{n_i} \sum_{k=(r-1)n_i+1}^{rn_i} Z_{N, \pi(k)}^2.
\eqa

Now consider the conditional expectation of $A$
\bqa
E(A | \vY) &=& \frac{1}{\tilde{N}}\sumi a \sumr {t_i} \frac{1}{n_i} \sum_{k=(r-1)n_i+1}^{rn_i} Z_{N, k}^2\\
%&=&  \frac{1}{\tilde{N}} \sumi a \sumr t \frac{1}{n_i} \sum_{k=(r-1)n_i+1}^{rn_i} Z_{N, k}^2\\
&\stackrel{P}{ \to}&  \sumi a \frac{1}{b_i}\sums {t_i} (\sigma_{i s}^2 + \mu_{i s}^2)
\eqa
as well as
\bqa
\Var(A| \vY) &=&  \Var \left(\sums {\tilde{N}} d_{N, s}^{(i)} Z_{N, \pi(s)}^2 | \vZ \right)\\
&=& \sums {\tilde{N}} \left(d_{N, s}^{(i)}- d_{N, \cdot}^{-(i)}\right)^2 \frac{1}{\tilde{N}-1}\sums {\tilde{N}} \left(Z_{N, s}^2 - \frac{1}{\tilde{N}-1} \sumr {\tilde{N}} Z_{N, r}^2 \right)^2,
\eqa
which converges to 0 in probability as above and since we have\\ $\frac{1}{\tilde{N}-1}\sums {\tilde{N}} \left(Z_{N, s}^2 - \frac{1}{\tilde{N}-1} \sumr {\tilde{N}} Z_{N, r}^2 \right)^2 = \mathcal{O}_P(1)$  because of the existence of fourth moments. 

Now, consider $r \neq s$. We have that
\bqa
E(A | \vY) &=& \frac{1}{n_i} \sumk {n_i}E(Y_{ikr}^{\pi} Y_{iks}^{\pi} | \vY) = E(Y_{111}^{\pi} Y_{112}^{\pi}| \vY) {\color{red}?}\\
&=&\frac{1}{(\tilde{N})!} \sum_{\pi \in \mathcal{S}_{\tilde{N}}} Z_{N, \pi(1)} Z_{N, \pi(2)}\\
&=& \frac{1}{\tilde{N}(\tilde{N}-1)} \sum_{i \neq j} Z_{N, i} Z_{N, j}
\eqa
Consider $E(Z_{N, i} Z_{N, j})$. There are two possibilities: If $Z_{N,i}$ and $Z_{N, j}$ stem from different random vectors (i.e., from different individuals) they are independent and we can write $E(Z_{N, i} Z_{N, j}) = E(Z_{N, i}) E( Z_{N, j})$. If they stem from the same individual, we cannot rewrite the expectation and we denote it by $\gamma_{i,j}:=E(Z_{N, i} Z_{N, j})\in (-\infty, \infty)$. For every fixed $i$ there are $(t_i-1)$ possible $j$'s such that $Z_{N, i}$ and $Z_{N, j}$ come from the same individual. 
This implies:
\bqan\label{proof:a}
E \left(\frac{1}{\tilde{N}(\tilde{N}-1)} \sum_{i \neq j} Z_{N, i} Z_{N, j} \right) &=& \frac{1}{\tilde{N}(\tilde{N}-1)} 
\sum_{(i,j)\in \Xi} E(Z_{N, i}) E(Z_{N, j})\\
&+& \frac{1}{\tilde{N}(\tilde{N}-1)} \sumi {\tilde{N}} 
\sum_{(i,j) \in \varLambda } \gamma_{i,j},\nonumber
\eqan
where the index sets are defined as $\Xi = \{(i,j,): i\neq j \text{ and } i,j \text{ stem from different subjects} \}$ and 
$\varLambda =\{(i,j): i\neq j \text{ and } i,j \text{ stem from the same subject} \}$.\\
Because of the Cauchy-Schwarz inequality and Condition (2) it holds that 
\bqa
\sup_{i,j} |\gamma_{i,j}| \leq 2 \sup_{i} E(Z_{N,i}^2) \leq C < \infty.
\eqa
Thus, it follows that 
$$
\frac{1}{\tilde{N}(\tilde{N}-1)} \sumi {\tilde{N}} \sum_{(i,j)\in \varLambda} \gamma_{i,j} \leq \frac{\tilde{N} (\max{t_i}-1)}{\tilde{N}(\tilde{N}-1)} C \to 0
$$ 
as $N \to \infty$. For the first summand on the right hand side in Equation \eqref{proof:a}, it holds that
\bqa
\frac{1}{\tilde{N}(\tilde{N}-1)} \sum_{(i,j)\in \Xi} E(Z_{N, i}) E(Z_{N, j}) &= & \frac{1}{\tilde{N}(\tilde{N}-1)} \sumi a \sums {t_i} \sumj a \sum_{r=1}^{t_j} \mu_{is} \mu_{j r }- o(1)\\
& \to& \left( \sumi a \frac{1}{b_i} \sums {t_i} \mu_{i s} \right)^2.
\eqa
To complete the proof it remains to show that 
$\Var\left(\frac{1}{\tilde{N}(\tilde{N}-1)} \sum_{i \neq j} Z_{N, i} Z_{N, j} \right)  \to 0$. Thus, 
\bqa 
\Var \left(\frac{1}{\tilde{N}(\tilde{N}-1)} \sum_{i \neq j} Z_{N, i} Z_{N, j} \right) &=& \frac{1}{(\tilde{N}(\tilde{N}-1))^2} \sum_{i_1 \neq j_1} \sum_{i_2 \neq j_2} \Cov(Z_{N, i_1} Z_{N, j_1}, Z_{N, i_2} Z_{N, j_2}).
\eqa
As above, we distinguish between the cases $(i,j)\in\Xi$ and $(i,j)\in\varLambda$. 
If $Z_{N, i_1} Z_{N, j_1}$ and $Z_{N, i_2} Z_{N, j_2}$ 
stem from different individuals it holds that $\Cov(Z_{N, i_1} Z_{N, j_1}, Z_{N, i_2} Z_{N, j_2}) = 0$ because of independence.
In all other cases  it holds that 
\bqa
\Cov(Z_{N, i_1} Z_{N, j_1}, Z_{N, i_2} Z_{N, j_2}) \leq 2 \sup_i E(Z_{N, i}^4) = \tilde{C} < \infty
\eqa
because of assumption (2) and the Cauchy-Schwarz inequality. \\
Furthermore, for every fixed $i_1$ and $j_1$ there are less than $5 (\max{t_i})^4$ possibilities for $Z_{N, i_2} Z_{N, j_2}$ to stem from the same individual(s) as $Z_{N, i_1} Z_{N, j_1}$, such that at least one of the sums 
cancels out and $\Var\left(\frac{1}{\tilde{N}(\tilde{N}-1)} \sum_{i \neq j} Z_{N, i} Z_{N, j} \right)  \to 0$ for all $i, j$ as $N \to \infty$. 

This implies that for $r \neq s$ 
\bqa
\frac{n_i-1}{n_i} \left(\vwhV_i^{\pi}\right)_{r, s} = A - B \stackrel{P}{\to} \left( \sumi a \frac{1}{b_i} \sums {t_i} \mu_{i s} \right)^2 -\left( \sumi a \frac{1}{b_i} \sums {t_i} \mu_{is} \right)^2 = 0
\eqa
and for $r = s$ we have $(\vwhV_i^\pi)_{r, s} \stackrel{P}{\to} \sigma^2$.
Altogether, this proves the desired result.
\qed \\
We are now able to prove Theorem \ref{theo}.\\
\proofit
Applying the continuous mapping theorem together with Lemma \ref{1} yields conditional convergence in distribution given $\vY$
\bqa
\vH \sqrt{N} (\volY_{\cdot}^{\pi}- \volY_{\ldots}) \stackrel{d}{\to} N(0, \sigma^2 \vH \vD \vH'),
\eqa
where $\vD :=  \diag(\kappa_1^{-1}\vI_{t_1}, \dots, \kappa_a^{-1}\vI_{t_a})$. Moreover, we have convergence in probability
\bqa
\vH \vwhSigma^\pi \vH' \stackrel{P}{\to} \sigma^2 \vH \vD \vH'
\eqa
by Lemma \ref{2}. Since $\det(\vwhV_i^\pi) > 0$ almost surely for $N$ large enough due to $\vSigma > 0$, 
the corresponding Moore-Penrose inverse converges as well in probability 
and hence another application of the continuous mapping theorem proves the result using Theorem 9.2.2 in \cite{Rao}.
\qed

\section{Other resampling approaches}\label{supp:OtherResampling}

\subsection{Nonparametric bootstrap approach}

Here, we consider a nonparametric bootstrap sample $\vY^* = (Y_{111}^*,\ldots,Y_{an_at}^*)$ 
drawn with replacement from the pooled observation vector 
$\vY = (Y_{111},\ldots,Y_{an_at})$. Therefore, given the observations, the bootstrap components are all independent 
with identical distribution which is given by the empirical distribution of $\vY^*$. 
The WTS of the bootstrap sample is given by
\bqa
Q_N^* = N (\volY_{\cdot}^*)'\vH'(\vH \vwhSigma^* \vH')^+\vH \volY_{\cdot}^*,
\eqa
where $\volY_{\cdot}^*$ is the vector of means of the bootstrap sample and $\vwhSigma^*$ denotes their covariance matrix. 

\begin{satz}\label{theo:NPBS}
	The distribution of $Q_N^{*}$ conditioned on the observed data $\vY$ weakly converges to the central $\chi^2_f$ distribution in probability, where $f=rank(\vH)$. 
	In particular, we have
	\bqan 
	\sup_{x \in \renu} \left| P_{\vmu}(Q_N^{*}\leq x | \vY) - P_{\vmu_0}(Q_N \leq x) \right| \to 0
	\eqan
	in probability for any underlying parameter $\vmu \in \renu^{at}$ and $\vmu_0 \in H_0(\vH)$.
\end{satz}

\proofit
The result follows analogously to the proof of Theorem 3.1 in the paper.%Konietschke et al.~(2015).
\qed

Note, that a nonparametric bootstrap version based on drawing with replacement from the observation vectors as in 
Konietschke et al. (2015) performed considerably worse than the parametric bootstrap approach described below and is therefore not reported here.

In addition, we have also studied a nonparametric bootstrap version of the ATS (although this is in general not asymptotically correct) given by
\bqa
F_N^* = \frac{N}{\tr(\vT\vwhSigma^*)} (\volY_{\cdot}^*)' \vT  \volY_{\cdot}^*.
\eqa
A corresponding permutation version of the ATS has not been considered since it is also asymptotically only an approximation.

\subsection{Parametric bootstrap approach}
We have also considered a parametric bootstrap approach as studied by, e.g.~Konietschke et al.~(2015). Here, the parametric bootstrap variables are generated as
\bqa
\vY^\star_i \stackrel{i.i.d.}{\sim} N(\textbf{0}, \vwhV_i), \hspace{0.2cm} 1 \leq i \leq a.
\eqa
The idea behind this approach is to obtain a more accurate finite sample approximation by mimicking the given covariance structure of the original data. 
We can again compute the WTS and ATS from the parametric bootstrap vectors as
\bqa
Q_N^\star = N (\volY_{\cdot}^\star)'\vH'(\vH \vwhSigma^\star \vH')^+\vH \volY_{\cdot}^\star,
\eqa
and
\bqa
F_N^\star = \frac{N}{\tr(\vT\vwhSigma^\star)} (\volY_{\cdot}^\star)' \vT  \volY_{\cdot}^\star,
\eqa
where $\volY_{\cdot}^\star$ is the vector of means of the parametric bootstrap sample and $\vwhSigma^\star$ denotes their empirical covariance matrix.

\begin{satz}\label{theo:PBS}
	The distribution of $Q_N^{\star}$ conditioned on the observed data $\vY$ weakly converges to the central $\chi^2_f$ distribution in probability, where $f=rank(\vH)$. In particular, we have
	\bqan
	\sup_{x \in \renu} \left| P_{\vmu}(Q_N^{\star}\leq x | \vY) - P_{\vmu_0}(Q_N \leq x) \right| \to 0
	\eqan
	in probability for any underlying parameters $\vmu,\vmu_0 \in \renu^{at}$ with $ \vH\vmu_0 = \vnull$.\\
	Furthermore, for the ATS of the parametric bootstrap sample it also holds that
	\bqan
	\sup_{x \in \renu} \left| P_{\vmu}(F_N^{\star}\leq x | \vY) - P_{\vmu_0}(F_N \leq x) \right| \to 0
	\eqan
	in probability for any underlying parameters $\vmu,\vmu_0 \in \renu^{at}$ with $ \vH\vmu_0 = \vnull$. Thus, 
	the conditional distribution of $F_N^\star$ always approximates the null distribution of $F_N$. 
\end{satz}

\proofit
The result for the WTS follows analogously to the proof of Theorem 3.2 in Konietschke et al.~(2015). 
For the parametric bootstrap version of the ATS the result is obtained by the multivariate Lindeberg-Feller Theorem, 
the Continuous Mapping Theorem and another application of Slutsky's Theorem. The details are left to the reader.
\qed

\subsection{Type-I error rates}

In the following, we present the results of the detailed simulation studies conducted as described in Section \ref{sim} of the paper. For comparison, the results of the permutation approach are also included. The results for the hypothesis of no time effect $T$ are presented in Tables \ref{table:supp_nT}, \ref{table:supp_lT} and  \ref{table:supp_eT} for the normal, log-normal and exponential distribution, respectively. The results for the hypothesis of no group $\times$ time interaction are in Tables \ref{table:supp_nGT}, \ref{table:supp_lGT} and \ref{table:supp_eGT}, respectively.
The parametric bootstrap approach is denoted by PBS, the nonparametric bootstrap by NPBS. 
The results are again compared to the asymptotic quantiles, i.e.~the $F(\hat{\nu},\infty)$-quantile for the ATS and the 
$\chi^2_f$-quantile for the WTS. A permutation version of the ATS has not been considered for the reasons stated above. 
The covariance settings and the number of simulated individuals are the same as described in Section \ref{sim}.

\newpage

\begin{table}[H]
	\footnotesize
	\centering
	\caption{Simulation results for the hypothesis of no time effect with normal distribution.}
	\label{table:supp_nT}
	\begin{tabular}{c|c|c|cc||cc}
		\multicolumn{7}{c}{normal distribution}\\ \hline \hline
		\multicolumn{3}{c|}{T} & \multicolumn{2}{c||}{$t=4$} & \multicolumn{2}{c}{$t=8$}\\ \hline
		Cov. Setting & $\vn$ & Method	& ATS & WTS  &  ATS &  WTS    \\ \hline
		\multirow{12}{*}{1} & \multirow{4}{*}{$\vn^{(1)}$}	&	Permutation & NA & 0.050 & NA & 0.050 \\ 
		& &	PBS & 0.041 & 0.052 & 0.034 & 0.059 \\ 
		& &		NPBS & 0.048 & 0.047 & 0.052 & 0.050 \\ 
		& &	asymptotic & 0.046 & 0.085 & 0.040 & 0.177 \\  \cline{2-7}
		& \multirow{4}{*}{$\vn^{(2)}$}	&		Permutation & NA & 0.048 & NA  & 0.052 \\ 
		& &		PBS & 0.041 & 0.050 & 0.034 & 0.060 \\ 
		& &	NPBS & 0.046 & 0.048 & 0.052 & 0.050 \\ 
		& &		asymptotic & 0.046 & 0.086 & 0.040 & 0.177 \\ \cline{2-7}
		& \multirow{4}{*}{$\vn^{(3)}$}	&		Permutation & NA & 0.051 & NA & 0.052 \\ 
		& &		PBS & 0.045 & 0.048 & 0.040 & 0.050 \\ 
		& &	NPBS & 0.051 & 0.051 & 0.050 & 0.052 \\ 
		& &		asymptotic & 0.050 & 0.078 & 0.043 & 0.135 \\ \cline{1-7}
		\multirow{12}{*}{2} & \multirow{4}{*}{$\vn^{(1)}$}	&	Permutation & NA & 0.050 & NA & 0.051 \\ 
		& &		PBS & 0.046 & 0.052 & 0.036 & 0.059 \\ 
		& &	NPBS & 0.056 & 0.051 & 0.054 & 0.050 \\ 
		& &		asymptotic & 0.051 & 0.085 & 0.042 & 0.177 \\ \cline{2-7}
		& \multirow{4}{*}{$\vn^{(2)}$}	&	Permutation & NA & 0.051 & NA & 0.052 \\ 
		& &	PBS & 0.046 & 0.052 & 0.036 & 0.060 \\ 
		& &		NPBS & 0.057 & 0.051 & 0.056 & 0.052 \\ 
		& &	asymptotic & 0.052 & 0.086 & 0.043 & 0.177 \\ \cline{2-7}
		& \multirow{4}{*}{$\vn^{(3)}$}	&		Permutation & NA & 0.051 & NA & 0.052 \\ 
		& &		PBS & 0.049 & 0.048 & 0.038 & 0.049 \\ 
		& &	NPBS & 0.059 & 0.049 & 0.054 & 0.051 \\ 
		& &		asymptotic & 0.053 & 0.077 & 0.041 & 0.135 \\ \cline{1-7}
		\multirow{12}{*}{3} & \multirow{4}{*}{$\vn^{(1)}$} &		Permutation & NA & 0.052 & NA & 0.062 \\ 
		& &		PBS & 0.041 & 0.052 & 0.040 & 0.064 \\ 
		& &	NPBS & 0.052 & 0.052 & 0.069 & 0.061 \\ 
		& &		asymptotic & 0.046 & 0.092 & 0.044 & 0.198 \\ \cline{2-7}
		& \multirow{4}{*}{$\vn^{(2)}$}	&		Permutation & NA & 0.045 & NA & 0.042 \\ 
		& &	PBS & 0.047 & 0.052 & 0.043 & 0.056 \\ 
		& &		NPBS & 0.056 & 0.043 & 0.075 & 0.042 \\ 
		& &	asymptotic & 0.051 & 0.080 & 0.048 & 0.155 \\ \cline{2-7}
		& \multirow{4}{*}{$\vn^{(3)}$}	&	Permutation & NA & 0.053 & NA  & 0.054 \\ 
		& &	PBS & 0.047 & 0.050 & 0.044 & 0.049 \\ 
		& &		NPBS & 0.058 & 0.051 & 0.073 & 0.052 \\ 
		& &	asymptotic & 0.051 & 0.078 & 0.048 & 0.136 \\ \cline{1-7}
		%		\multirow{12}{*}{4} & \multirow{4}{*}{$\vn^{(1)}$}&	Permutation & NA & 0.059 & NA  & 0.081 \\  
		%		& &	PBS & 0.040 & 0.053 & 0.030 & 0.070 \\  
		%		& &		NPBS & 0.048 & 0.059 & 0.052 & 0.082  \\ 
		%		& &	asymptotic & 0.048 & 0.100 & 0.036 & 0.231 \\ \cline{2-7}
		%		& \multirow{4}{*}{$\vn^{(2)}$}	&	Permutation & NA & 0.043 & NA & 0.035 \\ 
		%		& &		PBS & 0.044 & 0.050 & 0.038 & 0.054 \\ 
		%		& &	NPBS & 0.048 & 0.042 & 0.049 & 0.034  \\ 
		%		& &		asymptotic & 0.048 & 0.077 & 0.042 & 0.138 \\ \cline{2-7}
		%		& \multirow{4}{*}{$\vn^{(3)}$}	&	Permutation & NA & 0.053 & NA  & 0.057  \\ 
		%		& &	PBS &  0.044 & 0.049 & 0.037 & 0.051 \\ 
		%		& &		NPBS & 0.050 & 0.053 & 0.050 & 0.057 \\ 
		%		& &	asymptotic &  0.049 & 0.079 & 0.041 & 0.140 \\ 
		\hline
	\end{tabular}
\end{table}

%\newpage
%\thispagestyle{empty}

\begin{table}[H]
	\footnotesize
	\centering
	\caption{Simulation results for the hypothesis of no time effect with log-normal distribution.}
	\label{table:supp_lT}
	\begin{tabular}{c|c|c|cc||cc}
		\multicolumn{7}{c}{log-normal distribution}\\ \hline \hline
		\multicolumn{3}{c|}{T} & \multicolumn{2}{c||}{$t=4$} & \multicolumn{2}{c}{$t=8$}\\ \hline
		Cov. Setting & $\vn$ & Method	& ATS & WTS  &  ATS &  WTS    \\ \hline
		\multirow{12}{*}{1} & \multirow{4}{*}{$\vn^{(1)}$}	&	Permutation & NA & 0.051 & NA & 0.047 \\ 
		& &		PBS & 0.026 & 0.055 & 0.017 & 0.075 \\ 
		& &		NPBS & 0.051 & 0.050 & 0.047 & 0.048 \\ 
		& &	asymptotic & 0.032 & 0.094 & 0.021 & 0.198 \\  \cline{2-7}
		& \multirow{4}{*}{$\vn^{(2)}$}	 &			Permutation & NA & 0.052 & NA & 0.046 \\ 
		& &		PBS & 0.025 & 0.058 & 0.016 & 0.074 \\ 
		& &	NPBS & 0.048 & 0.051 & 0.054 & 0.046 \\ 
		& &		asymptotic & 0.031 & 0.090 & 0.020 & 0.198 \\  \cline{2-7}
		& \multirow{4}{*}{$\vn^{(3)}$}	 &			Permutation & NA & 0.051 & NA & 0.048 \\ 
		& &	PBS & 0.026 & 0.056 & 0.019 & 0.077 \\ 
		& &		NPBS & 0.052 & 0.050 & 0.051 & 0.050 \\ 
		& &	asymptotic & 0.031 & 0.089 & 0.021 & 0.186 \\ \hline
		\multirow{12}{*}{2} & \multirow{4}{*}{$\vn^{(1)}$}	&	Permutation & NA & 0.067 & NA & 0.053 \\ 
		& &		PBS & 0.035 & 0.072 & 0.018 & 0.084 \\ 
		& &	NPBS & 0.060 & 0.066 & 0.052 & 0.053 \\ 
		& &		asymptotic & 0.040 & 0.110 & 0.022 & 0.207 \\  \cline{2-7}
		& \multirow{4}{*}{$\vn^{(2)}$}	 &		Permutation & NA & 0.067 & NA & 0.051 \\ 
		& &	PBS & 0.034 & 0.073 & 0.018 & 0.082 \\ 
		& &		NPBS & 0.061 & 0.066 & 0.057 & 0.052 \\ 
		& &	asymptotic & 0.040 & 0.107 & 0.022 & 0.203 \\  \cline{2-7}
		& \multirow{4}{*}{$\vn^{(3)}$}	 &			Permutation & NA & 0.070 & NA & 0.057 \\ 
		& &		PBS & 0.037 & 0.072 & 0.021 & 0.080 \\ 
		& &	NPBS & 0.065 & 0.068 & 0.057 & 0.057 \\ 
		& &		asymptotic & 0.042 & 0.107 & 0.024 & 0.197 \\ \hline
		\multirow{12}{*}{3} & \multirow{4}{*}{$\vn^{(1)}$}	&	Permutation & NA & 0.057 & NA & 0.064 \\ 
		& &	PBS & 0.027 & 0.059 & 0.021 & 0.082 \\ 
		& &		NPBS & 0.054 & 0.058 & 0.063 & 0.064 \\ 
		& &	asymptotic & 0.033 & 0.101 & 0.024 & 0.221 \\  \cline{2-7}
		& \multirow{4}{*}{$\vn^{(2)}$}	 &			Permutation & NA & 0.053 & NA & 0.048 \\ 
		& &		PBS & 0.031 & 0.060 & 0.028 & 0.075 \\ 
		& &	NPBS & 0.057 & 0.053 & 0.079 & 0.047 \\ 
		& &		asymptotic & 0.037 & 0.090 & 0.033 & 0.190 \\  \cline{2-7}
		& \multirow{4}{*}{$\vn^{(3)}$}	 &		Permutation & NA & 0.057 & NA & 0.062 \\ 
		& &	PBS & 0.031 & 0.059 & 0.027 & 0.079 \\ 
		& &		NPBS & 0.057 & 0.054 & 0.075 & 0.062 \\ 
		& &	asymptotic & 0.036 & 0.092 & 0.031 & 0.191 \\ \hline
		%		\multirow{12}{*}{4} & \multirow{4}{*}{$\vn^{(1)}$}	&		Permutation & NA &  0.056 & NA  & 0.067  \\ 
		%		& &		PBS & 0.024 & 0.054 & 0.015 & 0.081 \\ 
		%		& &	NPBS &  0.051 & 0.056 & 0.048 & 0.068 \\ 
		%		& &		asymptotic & 0.029 & 0.103 & 0.018 & 0.239  \\  \cline{2-7}
		%		& \multirow{4}{*}{$\vn^{(2)}$}	 &		Permutation & NA & 0.048 & NA & 0.035  \\ 
		%		& &	PBS & 0.027 & 0.057 & 0.018 & 0.073 \\ 
		%		& &		NPBS &  0.047 & 0.048 & 0.053 & 0.036\\ 
		%		& &	asymptotic &  0.030 & 0.086 & 0.021 & 0.169 \\  \cline{2-7}
		%		& \multirow{4}{*}{$\vn^{(3)}$}	 &		Permutation & NA & 0.052 & NA & 0.053 \\ 
		%		& &		PBS &  0.027 & 0.057 & 0.018 & 0.076 \\ 
		%		& &	NPBS & 0.052 & 0.051 & 0.051 & 0.053 \\ 
		%		& &		asymptotic & 0.031 & 0.090 & 0.020 & 0.188 \\ 
		%		\hline
	\end{tabular}
\end{table}

%\newpage
%\thispagestyle{empty}

\begin{table}[H]
	\footnotesize
	\centering
	\caption{Simulation results for the hypothesis of no time effect with exponential distribution.}
	\label{table:supp_eT}
	\begin{tabular}{c|c|c|cc||cc}
		\multicolumn{7}{c}{exponential distribution}\\ \hline \hline
		\multicolumn{3}{c|}{T} & \multicolumn{2}{c||}{$t=4$} & \multicolumn{2}{c}{$t=8$}\\ \hline
		Cov. Setting & $\vn$ & Method	& ATS & WTS  &  ATS &  WTS    \\ \hline
		\multirow{12}{*}{1} & \multirow{4}{*}{$\vn^{(1)}$}	&		Permutation & NA & 0.048 & NA & 0.051 \\ 
		& &		PBS & 0.038 & 0.055 & 0.026 & 0.070 \\ 
		& &		NPBS & 0.052 & 0.049 & 0.052 & 0.051 \\ 
		& &	asymptotic & 0.045 & 0.090 & 0.034 & 0.194 \\ \cline{2-7}
		& \multirow{4}{*}{$\vn^{(2)}$}	 &	Permutation & NA & 0.053 & NA & 0.048 \\ 
		& &		PBS & 0.039 & 0.057 & 0.029 & 0.069 \\ 
		& &	NPBS & 0.053 & 0.053 & 0.050 & 0.048 \\ 
		& &	asymptotic & 0.046 & 0.096 & 0.032 & 0.191 \\ \cline{2-7}
		& \multirow{4}{*}{$\vn^{(3)}$}	 &		Permutation & NA & 0.054 & NA & 0.050 \\ 
		& &	PBS & 0.041 & 0.057 & 0.031 & 0.059 \\ 
		& &	NPBS & 0.053 & 0.054 & 0.049 & 0.050 \\ 
		& &	asymptotic & 0.046 & 0.086 & 0.034 & 0.151 \\ \hline
		\multirow{12}{*}{2} & \multirow{4}{*}{$\vn^{(1)}$}	&		Permutation & NA & 0.054 & NA & 0.052 \\ 
		& &	PBS & 0.040 & 0.059 & 0.029 & 0.070 \\ 
		& &	NPBS & 0.062 & 0.055 & 0.057 & 0.052 \\ 
		& &	asymptotic & 0.048 & 0.093 & 0.035 & 0.194 \\ \cline{2-7}
		& \multirow{4}{*}{$\vn^{(2)}$}	 &		Permutation & NA & 0.060 & NA & 0.051 \\ 
		& &	PBS & 0.044 & 0.064 & 0.029 & 0.074 \\ 
		& &	NPBS & 0.063 & 0.059 & 0.056 & 0.052 \\ 
		& &	asymptotic & 0.050 & 0.101 & 0.034 & 0.193 \\ \cline{2-7}
		& \multirow{4}{*}{$\vn^{(3)}$}	 &		Permutation & NA & 0.058 & NA  & 0.051 \\ 
		& &	PBS & 0.045 & 0.062 & 0.032 & 0.062 \\ 
		& &	NPBS & 0.060 & 0.058 & 0.052 & 0.051 \\ 
		& &	asymptotic & 0.050 & 0.088 & 0.036 & 0.154 \\ \hline
		\multirow{12}{*}{3} & \multirow{4}{*}{$\vn^{(1)}$}	&	Permutation & NA & 0.055 & NA & 0.066 \\ 
		& &	PBS & 0.041 & 0.055 & 0.034 & 0.074 \\ 
		& &	NPBS & 0.060 & 0.054 & 0.073 & 0.065 \\ 
		& &	asymptotic & 0.049 & 0.098 & 0.042 & 0.218 \\ \cline{2-7}
		& \multirow{4}{*}{$\vn^{(2)}$}	 &		Permutation & NA & 0.049 & NA & 0.045 \\ 
		& &	PBS & 0.045 & 0.058 & 0.039 & 0.067 \\ 
		& &	NPBS & 0.062 & 0.049 & 0.078 & 0.044 \\ 
		& &		asymptotic & 0.050 & 0.090 & 0.046 & 0.173 \\ \cline{2-7}
		& \multirow{4}{*}{$\vn^{(3)}$}	 &		Permutation & NA & 0.055 & NA & 0.056 \\ 
		& &	PBS & 0.045 & 0.058 & 0.038 & 0.060 \\ 
		& &	NPBS & 0.062 & 0.055 & 0.072 & 0.056 \\ 
		& &	asymptotic & 0.050 & 0.087 & 0.042 & 0.153 \\ \hline
		%		\multirow{12}{*}{4} & \multirow{4}{*}{$\vn^{(1)}$}	&		Permutation & NA   & 0.059 & NA & 0.078  \\ 
		%		& &	PBS &  0.035 & 0.056 & 0.023 & 0.076 \\ 
		%		& &	NPBS & 0.056 & 0.060 & 0.053 & 0.079  \\ 
		%		& &	asymptotic & 0.044 & 0.104 & 0.030 & 0.250 \\ \cline{2-7}
		%		& \multirow{4}{*}{$\vn^{(2)}$}	 &		Permutation & NA & 0.045 & NA & 0.035 \\ 
		%		& &	PBS & 0.042 & 0.056 & 0.032 & 0.065 \\ 
		%		& &	NPBS &  0.053 & 0.046 & 0.051 & 0.034 \\ 
		%		& &	asymptotic & 0.047 & 0.086 & 0.036 & 0.153 \\ \cline{2-7}
		%		& \multirow{4}{*}{$\vn^{(3)}$}	 &		Permutation & NA & 0.054 & NA  & 0.054 \\ 
		%		& &	PBS & 0.041 & 0.057 & 0.031 & 0.062  \\ 
		%		& &	NPBS &  0.054 & 0.055 & 0.051 & 0.053  \\ 
		%		& &	asymptotic & 0.045 & 0.086 & 0.035 & 0.158 \\ 
		%		\hline
	\end{tabular}
\end{table}

%\newpage
%\thispagestyle{empty}

\begin{table}[H]
	\footnotesize
	\centering
	\caption{Simulation results for the hypothesis of no group $\times$ time interaction with normal distribution.}
	\label{table:supp_nGT}
	\begin{tabular}{c|c|c|cc||cc}
		\multicolumn{7}{c}{normal distribution}\\ \hline \hline
		\multicolumn{3}{c|}{GT} & \multicolumn{2}{c||}{$t=4$} & \multicolumn{2}{c}{$t=8$}\\ \hline
		Cov. Setting & $\vn$ & Method	& ATS & WTS  &  ATS &  WTS    \\ \hline
		\multirow{12}{*}{1} & \multirow{4}{*}{$\vn^{(1)}$}	&	Permutation & NA & 0.046 & NA & 0.051 \\ 
		& &		PBS & 0.039 & 0.051 & 0.025 & 0.077 \\ 
		& &	NPBS & 0.049 & 0.046 & 0.052 & 0.051 \\ 
		& &		asymptotic & 0.049 & 0.135 & 0.033 & 0.432 \\  \cline{2-7}
		& \multirow{4}{*}{$\vn^{(2)}$}	 &		Permutation & NA & 0.052 & NA & 0.050 \\ 
		& &	PBS & 0.042 & 0.056 & 0.026 & 0.075 \\ 
		& &		NPBS & 0.052 & 0.051 & 0.051 & 0.049 \\ 
		& &	asymptotic & 0.053 & 0.142 & 0.034 & 0.433 \\  \cline{2-7}
		& \multirow{4}{*}{$\vn^{(3)}$}	&	Permutation & NA & 0.049 & NA & 0.051 \\ 
		& &		PBS & 0.041 & 0.049 & 0.032 & 0.046 \\ 
		& &	NPBS & 0.050 & 0.050 & 0.054 & 0.050 \\ 
		& &		asymptotic & 0.048 & 0.126 & 0.039 & 0.366 \\  \cline{1-7}
		\multirow{12}{*}{2} & \multirow{4}{*}{$\vn^{(1)}$}	&	Permutation & NA & 0.050 & NA & 0.052 \\ 
		& &		PBS & 0.045 & 0.054 & 0.030 & 0.076 \\ 
		& &		NPBS & 0.060 & 0.050 & 0.055 & 0.053 \\ 
		& &	asymptotic & 0.053 & 0.132 & 0.038 & 0.429 \\  \cline{2-7}
		& \multirow{4}{*}{$\vn^{(2)}$}	 &		Permutation & NA & 0.054 & NA & 0.050 \\ 
		& &		PBS & 0.044 & 0.056 & 0.029 & 0.072 \\ 
		& &		NPBS & 0.059 & 0.053 & 0.057 & 0.051 \\ 
		& &	asymptotic & 0.053 & 0.141 & 0.038 & 0.431 \\  \cline{2-7}
		& \multirow{4}{*}{$\vn^{(3)}$}	&		Permutation & NA & 0.052 & NA & 0.050 \\ 
		& &		PBS & 0.044 & 0.049 & 0.034 & 0.046 \\ 
		& &		NPBS & 0.059 & 0.052 & 0.060 & 0.050 \\ 
		& &	asymptotic & 0.050 & 0.122 & 0.040 & 0.366 \\  \cline{1-7}
		\multirow{12}{*}{3} & \multirow{4}{*}{$\vn^{(1)}$}	&	Permutation & NA & 0.050 & NA & 0.065 \\ 
		& &		PBS & 0.043 & 0.051 & 0.033 & 0.082 \\ 
		& &	NPBS & 0.061 & 0.049 & 0.075 & 0.069 \\ 
		& &		asymptotic & 0.054 & 0.141 & 0.040 & 0.465 \\  \cline{2-7}
		& \multirow{4}{*}{$\vn^{(2)}$}	&	Permutation & NA & 0.045 & NA & 0.037 \\ 
		& &	PBS & 0.046 & 0.054 & 0.040 & 0.069 \\ 
		& &		NPBS & 0.057 & 0.047 & 0.078 & 0.037 \\ 
		& &	asymptotic & 0.053 & 0.135 & 0.049 & 0.393 \\  \cline{2-7}
		& \multirow{4}{*}{$\vn^{(3)}$}	&	Permutation & NA & 0.049 &  NA & 0.053 \\ 
		& &		PBS & 0.043 & 0.048 & 0.038 & 0.047 \\ 
		& &	NPBS & 0.064 & 0.050 & 0.077 & 0.051 \\ 
		& &		asymptotic & 0.051 & 0.126 & 0.045 & 0.363 \\  \cline{1-7}
		%		\multirow{12}{*}{4} & \multirow{4}{*}{$\vn^{(1)}$}	&	Permutation & NA &  0.055 & NA & 0.092 \\ 
		%		& &	PBS &  0.040 & 0.052 & 0.024 & 0.090 \\ 
		%		& &		NPBS & 0.061 & 0.056 & 0.062 & 0.091 \\ 
		%		& &	asymptotic & 0.051 & 0.146 & 0.033 & 0.503\\  \cline{2-7}
		%		& \multirow{4}{*}{$\vn^{(2)}$}	&	Permutation & NA & 0.044 & NA & 0.027 \\ 
		%		& &	PBS & 0.042 & 0.053 & 0.027 & 0.063  \\ 
		%		& &		NPBS & 0.045 & 0.044 & 0.041 & 0.026 \\ 
		%		& &	asymptotic &  0.050 & 0.129 & 0.033 & 0.358 \\  \cline{2-7}
		%		& \multirow{4}{*}{$\vn^{(3)}$}	&		Permutation & NA & 0.052 & NA & 0.055 \\ 
		%		& &		PBS &  0.043 & 0.050 & 0.032 & 0.046 \\ 
		%		& &	NPBS & 0.055 & 0.051 & 0.056 & 0.053 \\ 
		%		& &		asymptotic &  0.049 & 0.127 & 0.038 & 0.368 \\ 
		%		\hline
	\end{tabular}
\end{table}

%\newpage
%\thispagestyle{empty}

\begin{table}[H]
	\footnotesize
	\centering
	\caption{Simulation results for the hypothesis of no group $\times$ time interaction with log-normal distribution.}
	\label{table:supp_lGT}
	\begin{tabular}{c|c|c|cc||cc}
		\multicolumn{7}{c}{log-normal distribution}\\ \hline \hline
		\multicolumn{3}{c|}{GT} & \multicolumn{2}{c||}{$t=4$} & \multicolumn{2}{c}{$t=8$}\\ \hline
		Cov. Setting & $\vn$ & Method	& ATS & WTS  &  ATS &  WTS    \\ \hline
		\multirow{12}{*}{1} & \multirow{4}{*}{$\vn^{(1)}$}	&	Permutation & NA & 0.047 & NA & 0.053 \\ 
		& &		PBS & 0.019 & 0.040 & 0.009 & 0.061 \\ 
		& &		NPBS & 0.048 & 0.047 & 0.048 & 0.052 \\ 
		& &		asymptotic & 0.024 & 0.121 & 0.012 & 0.426 \\ \cline{2-7}
		& \multirow{4}{*}{$\vn^{(2)}$}	 &		Permutation & NA & 0.053 & NA & 0.051 \\ 
		& &		PBS & 0.017 & 0.044 & 0.009 & 0.055 \\ 
		& &	NPBS & 0.048 & 0.053 & 0.050 & 0.048 \\ 
		& &		asymptotic & 0.022 & 0.128 & 0.013 & 0.431 \\ \cline{2-7}
		& \multirow{4}{*}{$\vn^{(3)}$}	 &	Permutation & NA & 0.048 & NA & 0.051 \\ 
		& &		PBS & 0.018 & 0.037 & 0.010 & 0.042 \\ 
		& &		NPBS & 0.048 & 0.046 & 0.047 & 0.051 \\ 
		& &		asymptotic & 0.024 & 0.118 & 0.012 & 0.406 \\ \hline
		\multirow{12}{*}{2} & \multirow{4}{*}{$\vn^{(1)}$}	&	Permutation & NA & 0.051 & NA & 0.054 \\ 
		& &		PBS & 0.019 & 0.044 & 0.010 & 0.062 \\ 
		& &	NPBS & 0.056 & 0.051 & 0.052 & 0.054 \\ 
		& &		asymptotic & 0.025 & 0.129 & 0.014 & 0.427 \\ \cline{2-7}
		& \multirow{4}{*}{$\vn^{(2)}$}	 &	Permutation & NA & 0.054 & NA & 0.052 \\ 
		& &		PBS & 0.019 & 0.044 & 0.011 & 0.056 \\ 
		& &	NPBS & 0.056 & 0.053 & 0.052 & 0.051 \\ 
		& &		asymptotic & 0.026 & 0.130 & 0.013 & 0.432 \\ \cline{2-7}
		& \multirow{4}{*}{$\vn^{(3)}$}	 &		Permutation & NA & 0.050 & NA & 0.052 \\ 
		& &	PBS & 0.018 & 0.038 & 0.010 & 0.042 \\ 
		& &		NPBS & 0.056 & 0.049 & 0.053 & 0.052 \\ 
		& &		asymptotic & 0.023 & 0.120 & 0.013 & 0.403 \\ \hline
		\multirow{12}{*}{3} & \multirow{4}{*}{$\vn^{(1)}$}	&		Permutation & NA & 0.050 & NA & 0.062 \\ 
		& &	PBS & 0.022 & 0.042 & 0.014 & 0.067 \\ 
		& &		NPBS & 0.053 & 0.050 & 0.068 & 0.060 \\ 
		& &	asymptotic & 0.029 & 0.133 & 0.020 & 0.457 \\ \cline{2-7}
		& \multirow{4}{*}{$\vn^{(2)}$}	 &		Permutation & NA & 0.045 & NA & 0.036 \\ 
		& &		PBS & 0.022 & 0.043 & 0.020 & 0.053 \\ 
		& &		NPBS & 0.055 & 0.046 & 0.076 & 0.035 \\ 
		& &		asymptotic & 0.028 & 0.121 & 0.024 & 0.399 \\ \cline{2-7}
		& \multirow{4}{*}{$\vn^{(3)}$}	 &		Permutation & NA & 0.049 & NA & 0.053 \\ 
		& &	PBS & 0.023 & 0.037 & 0.014 & 0.043 \\ 
		& &		NPBS & 0.059 & 0.046 & 0.071 & 0.054 \\ 
		& &		asymptotic & 0.028 & 0.122 & 0.020 & 0.408 \\ \hline
		%		\multirow{12}{*}{4} & \multirow{4}{*}{$\vn^{(1)}$}	&		Permutation & NA &  0.057 & NA  & 0.093\\ 
		%		& &	PBS & 0.018 & 0.043 & 0.009 & 0.075  \\ 
		%		& &		NPBS & 0.054 & 0.056 & 0.053 & 0.092 \\ 
		%		& &		asymptotic & 0.024 & 0.137 & 0.012 & 0.497 \\ \cline{2-7}
		%		& \multirow{4}{*}{$\vn^{(2)}$}	 &	Permutation & NA & 0.045 & NA  & 0.032 \\ 
		%		& &	PBS & 0.019 & 0.042 & 0.011 & 0.048 \\ 
		%		& &	NPBS & 0.045 & 0.045 & 0.045 & 0.031 \\ 
		%		& &	asymptotic & 0.023 & 0.114 & 0.013 & 0.369 \\ \cline{2-7}
		%		& \multirow{4}{*}{$\vn^{(3)}$}	 &		Permutation & NA & 0.051 & NA  & 0.059 \\ 
		%		& &		PBS & 0.018 & 0.037 & 0.008 & 0.046\\ 
		%		& &	NPBS & 0.049 & 0.050 & 0.048 & 0.057 \\ 
		%		& &		asymptotic & 0.024 & 0.121 & 0.011 & 0.414 \\ 
		%		\hline
	\end{tabular}
\end{table}

%\newpage
%\thispagestyle{empty}

\begin{table}[H]
	\footnotesize
	\centering
	\caption{Simulation results for the hypothesis of no group $\times$ time interaction with exponential distribution.}
	\label{table:supp_eGT}
	\begin{tabular}{c|c|c|cc||cc}
		\multicolumn{7}{c}{exponential distribution}\\ \hline \hline
		\multicolumn{3}{c|}{GT} & \multicolumn{2}{c||}{$t=4$} & \multicolumn{2}{c}{$t=8$}\\ \hline
		Cov. Setting & $\vn$ & Method	& ATS & WTS  &  ATS &  WTS    \\ \hline
		\multirow{12}{*}{1} & \multirow{4}{*}{$\vn^{(1)}$}	&		Permutation & NA & 0.054 &  NA & 0.054 \\ 
		& &	PBS & 0.036 & 0.057 & 0.018 & 0.076 \\ 
		& &	NPBS & 0.055 & 0.055 & 0.049 & 0.055 \\ 
		& &	asymptotic & 0.043 & 0.146 & 0.024 & 0.442 \\ \cline{2-7}
		& \multirow{4}{*}{$\vn^{(2)}$}	 &	Permutation & NA & 0.054 & NA & 0.050 \\ 
		& &		PBS & 0.030 & 0.057 & 0.019 & 0.072 \\ 
		& &		NPBS & 0.051 & 0.054 & 0.048 & 0.050 \\ 
		& &	asymptotic & 0.041 & 0.148 & 0.024 & 0.443 \\ \cline{2-7}
		& \multirow{4}{*}{$\vn^{(3)}$}	 &		Permutation & NA & 0.047 & NA & 0.054 \\ 
		& &	PBS & 0.029 & 0.043 & 0.023 & 0.052 \\ 
		& &	NPBS & 0.047 & 0.046 & 0.054 & 0.054 \\ 
		& &	asymptotic & 0.036 & 0.122 & 0.028 & 0.397 \\ \hline
		\multirow{12}{*}{2} & \multirow{4}{*}{$\vn^{(1)}$}	&	Permutation & NA & 0.059 & NA & 0.057 \\ 
		& &		PBS & 0.040 & 0.061 & 0.019 & 0.077 \\ 
		& &	NPBS & 0.065 & 0.061 & 0.054 & 0.056 \\ 
		& &	asymptotic & 0.048 & 0.151 & 0.027 & 0.444 \\ \cline{2-7}
		& \multirow{4}{*}{$\vn^{(2)}$}	 &		Permutation & NA & 0.059 & NA & 0.052 \\ 
		& &	PBS & 0.035 & 0.060 & 0.019 & 0.072 \\ 
		& &	NPBS & 0.058 & 0.059 & 0.050 & 0.051 \\ 
		& &	asymptotic & 0.042 & 0.153 & 0.025 & 0.448 \\ \cline{2-7}
		& \multirow{4}{*}{$\vn^{(3)}$}	 &		Permutation & NA & 0.048 & NA & 0.055 \\ 
		& &	PBS & 0.028 & 0.042 & 0.024 & 0.051 \\ 
		& &	NPBS & 0.051 & 0.048 & 0.058 & 0.054 \\ 
		& &	asymptotic & 0.034 & 0.121 & 0.029 & 0.397 \\ \hline
		\multirow{12}{*}{3} & \multirow{4}{*}{$\vn^{(1)}$}	&	Permutation & NA & 0.061 & NA & 0.068 \\ 
		& &	PBS & 0.039 & 0.059 & 0.024 & 0.083 \\ 
		& &	NPBS & 0.067 & 0.060 & 0.068 & 0.069 \\ 
		& &	asymptotic & 0.047 & 0.155 & 0.032 & 0.473 \\ \cline{2-7}
		& \multirow{4}{*}{$\vn^{(2)}$}	 &		Permutation & NA & 0.049 & NA & 0.037 \\ 
		& &		PBS & 0.038 & 0.056 & 0.033 & 0.062 \\ 
		& &	NPBS & 0.056 & 0.050 & 0.078 & 0.036 \\ 
		& &	asymptotic & 0.043 & 0.140 & 0.042 & 0.406 \\ \cline{2-7}
		& \multirow{4}{*}{$\vn^{(3)}$}	 &		Permutation & NA & 0.047 & NA & 0.058 \\ 
		& &	PBS & 0.031 & 0.043 & 0.033 & 0.052 \\ 
		& &	NPBS & 0.055 & 0.047 & 0.080 & 0.057 \\ 
		& &	asymptotic & 0.037 & 0.122 & 0.041 & 0.402 \\ \hline
		%		\multirow{12}{*}{4} & \multirow{4}{*}{$\vn^{(1)}$}	&		Permutation & NA &  0.066 & NA & 0.100  \\ 
		%		& &	PBS & 0.034 & 0.060 & 0.016 & 0.096  \\ 
		%		& &	NPBS &  0.064 & 0.065 & 0.060 & 0.099\\ 
		%		& &	asymptotic &  0.042 & 0.167 & 0.023 & 0.514\\ \cline{2-7}
		%		& \multirow{4}{*}{$\vn^{(2)}$}	 &		Permutation & NA & 0.046 &  NA & 0.028  \\ 
		%		& &	PBS & 0.034 & 0.054 & 0.021 & 0.057 \\ 
		%		& &	NPBS & 0.046 & 0.045 & 0.041 & 0.028 \\ 
		%		& &	asymptotic & 0.041 & 0.131 & 0.027 & 0.373  \\ \cline{2-7}
		%		& \multirow{4}{*}{$\vn^{(3)}$}	 &		Permutation & NA  &  0.051 & NA & 0.060\\ 
		%		& &	PBS & 0.030 & 0.045 & 0.023 & 0.056 \\ 
		%		& &	NPBS &  0.049 & 0.049 & 0.058 & 0.060  \\ 
		%		& &	asymptotic & 0.035 & 0.123 & 0.030 & 0.408 \\ 
		%	\hline
	\end{tabular}
\end{table}

\section{Additional simulation results: Quality of the approximation} \label{supp:AdditionalResults}

\begin{figure}[H]
	\centering
	\includegraphics[width=0.9\textwidth]{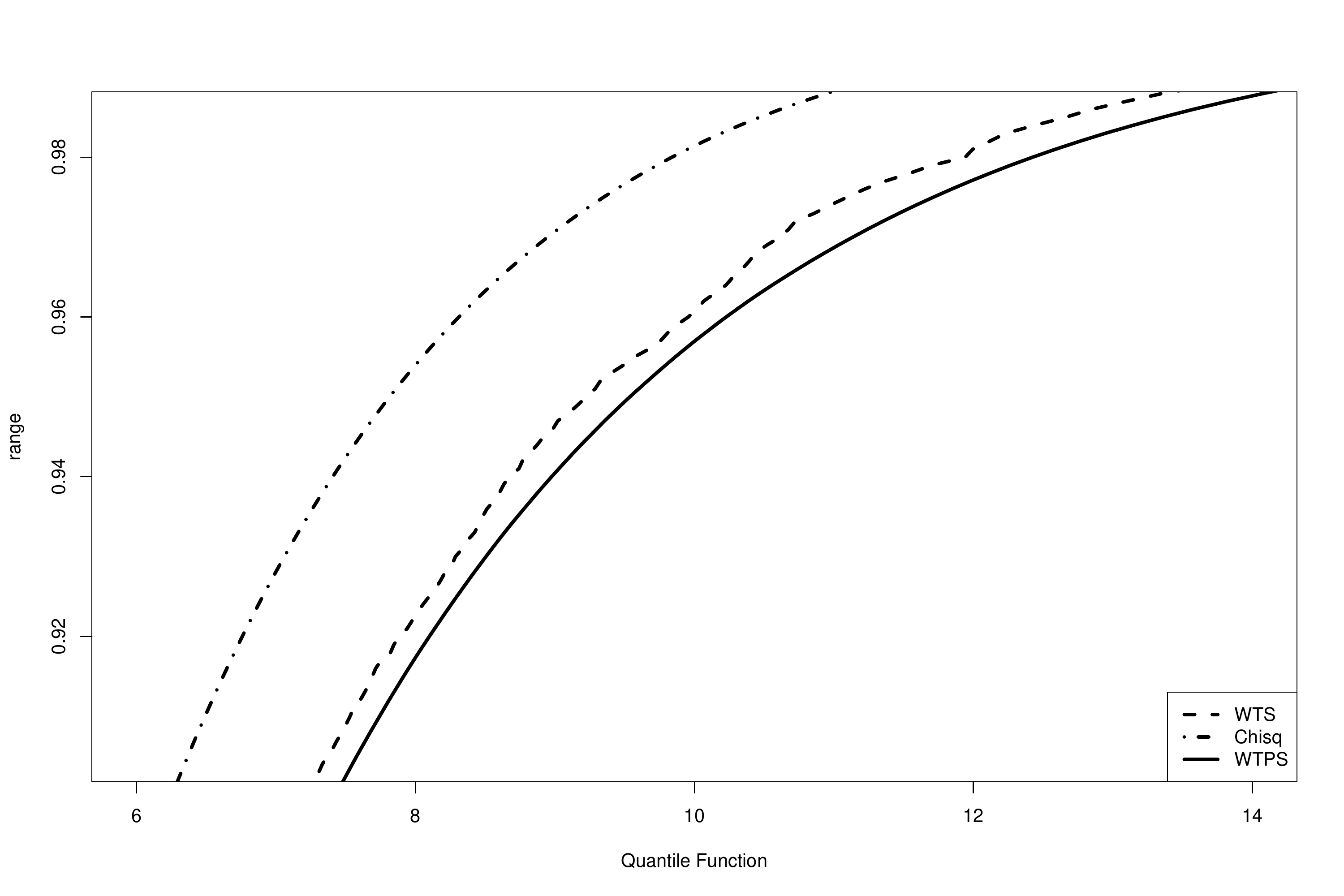}
	\caption{\it Quantile funtions of the WTS, WTPS and the corresponding $\chi^2$-distribution in the simulation setting with normally distributed data, $t=4$, covariance matrix setting 3, $\vn^{(2)}$ and under the null hypothesis of no time effect.}
	\label{Fig.QF_normal}
\end{figure}

Recall that we defined
$$
KQS = \sup_{0.9 \leq t \leq 0.99} | F_N^{-1}(t) - F^{-1}(t)|
$$
as well as
$$
KQS^\pi = \sup_{0.9 \leq t \leq 0.99} | F_N^{-1}(t) - (F_N^\pi)^{-1}(t)|
$$ for the distances between the quantile functions of the WTS ($F_N^{-1}$) and the $\chi^2$-distribution($F^{-1}$) and the WTPS($(F_N^\pi)^{-1}$), respectively. The results for all simulation settings described in the paper are presented in Tables \ref{KQS1} and \ref{KQS2}.
A plot of one exemplarily chosen scenario is in Figure \ref{Fig.QF_normal}.

\newpage

\begin{table}[H]
	\footnotesize
	\centering
	\caption{Simulation results for the distances between the quantile functions for the hypothesis of no time effect.}
	\label{KQS1}
	\begin{tabular}{c|c|cc||cc}
		\multicolumn{6}{c}{normal distribution}\\ \hline \hline
		\multicolumn{2}{c|}{T} & \multicolumn{2}{c||}{$t=4$} & \multicolumn{2}{c}{$t=8$}\\ \hline
		Cov. Setting &	& $\chi^2$ & WTPS & $\chi^2$ & WTPS    \\ \hline
		
		\multirow{3}{*}{1} & $\vn^{(1)}$	& 3.683 & 0.411 & 10.548 & 0.381 \\ \cline{2-6}
		& $\vn^{(2)}$  & 3.299 & 0.310 & 11.393 & 0.654 \\ \cline{2-6}
		& $\vn^{(3)}$ & 2.198 & 0.213 & 7.771 & 0.281 \\ \hline\hline
		\multirow{3}{*}{2} & $\vn^{(1)}$	 & 3.620 & 0.564 & 10.494 & 0.286 \\ \cline{2-6}
		& $\vn^{(2)}$  & 3.378 & 0.227 & 11.604 & 0.993 \\ \cline{2-6}
		& $\vn^{(3)}$ & 1.991 & 0.226 & 7.646 & 0.297 \\ \hline\hline
		\multirow{3}{*}{3} & $\vn^{(1)}$	 & 4.451 & 1.186 & 12.515 & 2.086 \\ \cline{2-6}
		& $\vn^{(2)}$  & 2.599 & 0.731 & 9.564 & 1.486 \\ \cline{2-6}
		& $\vn^{(3)}$ & 2.264 & 0.105 & 7.571 & 0.344 \\ \hline\hline
		%		\multirow{3}{*}{4} & $\vn^{(1)}$	 & 5.114 & 1.834 & 15.426 & 4.838 \\ \cline{2-6}
		%		& $\vn^{(2)}$  & 2.188 & 1.148 & 8.305 & 2.800 \\ \cline{2-6}
		%		& $\vn^{(3)}$  & 2.248 & 0.291 & 8.146 & 0.923 \\  \hline
		%		\hline
		
		\multicolumn{6}{c}{log-normal distribution}\\ \hline \hline
		\multicolumn{2}{c|}{T} & \multicolumn{2}{c||}{$t=4$} & \multicolumn{2}{c}{$t=8$}\\ \hline
		Cov. Setting &	& $\chi^2$ & WTPS & $\chi^2$ & WTPS    \\ \hline
		\multirow{3}{*}{1} & $\vn^{(1)}$	 & 3.097 & 0.532 & 12.399 & 1.044 \\  \cline{2-6}
		& $\vn^{(2)}$ & 3.960 & 0.386 & 14.293 & 0.998 \\  \cline{2-6}
		& $\vn^{(3)}$ & 3.087 & 0.363 & 12.768 & 0.421 \\ \hline\hline
		\multirow{3}{*}{2} & $\vn^{(1)}$	 & 5.645 & 2.258 & 14.165 & 1.105 \\  \cline{2-6}
		& $\vn^{(2)}$ & 6.045 & 2.656 & 15.484 & 2.617 \\  \cline{2-6}
		& $\vn^{(3)}$ & 5.062 & 1.891 & 14.239 & 1.815 \\ \hline\hline
		\multirow{3}{*}{3} & $\vn^{(1)}$	 & 3.977 & 0.517 & 14.547 & 2.299 \\  \cline{2-6}
		& $\vn^{(2)}$ & 4.000 & 0.526 & 12.740 & 0.610 \\  \cline{2-6}
		& $\vn^{(3)}$ & 3.561 & 0.643 & 14.238 & 3.203 \\ \hline\hline
		%		\multirow{3}{*}{4} & $\vn^{(1)}$	 & 3.898 & 0.465 & 16.356 & 3.307 \\  \cline{2-6}
		%		& $\vn^{(2)}$ & 3.317 & 0.468 & 11.033 & 2.288 \\  \cline{2-6}
		%		& $\vn^{(3)}$ & 3.165 & 0.175 & 13.693 & 1.411 \\  \hline
		%		\hline
		%	\end{tabular}
		%\end{table}
		%
		%\begin{table}[H]
		%	\centering
		%	\begin{tabular}{c|c|ccc||ccc}
		\multicolumn{6}{c}{exponential distribution}\\ \hline \hline
		\multicolumn{2}{c|}{T} & \multicolumn{2}{c||}{$t=4$} & \multicolumn{2}{c}{$t=8$}\\ \hline
		Cov. Setting &	& $\chi^2$ & WTPS & $\chi^2$ & WTPS    \\ \hline
		\multirow{3}{*}{1} & $\vn^{(1)}$ &  3.617 & 0.283 & 11.750 & 1.054 \\  \cline{2-6}
		& $\vn^{(2)}$  & 4.245 & 0.491 & 12.098 & 0.948 \\  \cline{2-6}
		& $\vn^{(3)}$ & 2.906 & 0.382 & 9.685 & 0.885 \\ \hline\hline
		\multirow{3}{*}{2} & $\vn^{(1)}$ & 4.761 & 1.262 & 11.704 & 0.628 \\  \cline{2-6}
		& $\vn^{(2)}$  & 4.961 & 1.366 & 12.201 & 0.724 \\  \cline{2-6}
		& $\vn^{(3)}$ & 3.833 & 0.915 & 10.226 & 0.286 \\ \hline\hline
		\multirow{3}{*}{3} & $\vn^{(1)}$ & 4.567 & 0.969 & 13.840 & 2.089 \\  \cline{2-6}
		& $\vn^{(2)}$  & 3.700 & 0.290 & 10.504 & 1.729 \\  \cline{2-6}
		& $\vn^{(3)}$ & 3.179 & 0.343 & 10.093 & 0.601 \\ \hline\hline
		%		\multirow{3}{*}{4} & $\vn^{(1)}$ & 4.835 & 1.290 & 15.789 & 3.577 \\  \cline{2-6}
		%		& $\vn^{(2)}$ & 3.105 & 0.729 & 9.672 & 3.118 \\  \cline{2-6}
		%		& $\vn^{(3)}$  & 3.204 & 0.244 & 10.044 & 0.362 \\  \hline
		%		\hline
	\end{tabular}
\end{table}

\newpage

\begin{table}[H]
	\centering
	\footnotesize
	\caption{Simulation results for the distances between the quantile functions for the hypothesis of no interaction.}
	\label{KQS2}
	\begin{tabular}{c|c|cc||cc}
		\multicolumn{6}{c}{normal distribution}\\ \hline \hline
		\multicolumn{2}{c|}{GT} & \multicolumn{2}{c||}{$t=4$} & \multicolumn{2}{c}{$t=8$}\\ \hline
		Cov. Setting &	& $\chi^2$ & WTPS & $\chi^2$ & WTPS    \\ \hline
		
		\multirow{3}{*}{1} & n1	 & 9.151 & 0.420 & 40.954 & 1.135 \\ \cline{2-6}
		& n2 & 8.872 & 0.573 & 41.179 & 1.451 \\ \cline{2-6}
		& n3 & 7.789 & 0.711 & 30.617 & 0.905 \\ \hline\hline
		\multirow{3}{*}{2} & n1	 & 8.648 & 0.582 & 42.023 & 1.804 \\\cline{2-6} 
		& n2 & 8.727 & 0.497 & 41.980 & 1.928 \\ \cline{2-6}
		& n3 & 7.951 & 1.166 & 30.031 & 0.956 \\ \hline\hline
		\multirow{3}{*}{3} & n1	 & 10.280 & 1.108 & 48.106 & 7.618 \\ \cline{2-6}
		& n2 & 7.700 & 1.463 & 36.252 & 4.470 \\ \cline{2-6}
		& n3 & 7.579 & 0.604 & 31.374 & 1.461 \\ \hline\hline
		%		\multirow{3}{*}{4} & n1	 & 11.539 & 2.610 & 58.753 & 18.433 \\ \cline{2-6}
		%		& n2 & 6.806 & 2.300 & 31.069 & 9.305 \\ \cline{2-6}
		%		& n3 & 7.536 & 0.866 & 31.702 & 2.061 \\  \hline
		%		\hline
		
		\multicolumn{6}{c}{log-normal distribution}\\ \hline \hline
		\multicolumn{2}{c|}{GT} & \multicolumn{2}{c||}{$t=4$} & \multicolumn{2}{c}{$t=8$}\\ \hline
		Cov. Setting &	& $\chi^2$ & WTPS & $\chi^2$ & WTPS    \\ \hline
		\multirow{3}{*}{1} & n1	  & 5.292 & 0.610 & 30.474 & 1.058 \\ \cline{2-6}
		& n2   & 5.952 & 0.360 & 30.014 & 1.671 \\ \cline{2-6}
		& n3   & 5.767 & 0.467 & 27.019 & 1.024 \\ \hline\hline
		\multirow{3}{*}{2} & n1 & 5.340 & 0.524 & 30.986 & 0.770 \\ \cline{2-6}
		& n2 & 6.307 & 0.812 & 29.960 & 1.329 \\ \cline{2-6}
		& n3   & 5.826 & 0.674 & 27.346 & 0.558 \\ \hline\hline
		\multirow{3}{*}{3} & n1 & 6.425 & 0.298 & 34.755 & 2.106 \\ \cline{2-6}
		& n2 & 5.124 & 1.408 & 26.657 & 6.691 \\ \cline{2-6}
		& n3   & 5.561 & 0.182 & 27.517 & 1.363 \\ \hline\hline
		%		\multirow{3}{*}{4} & n1 & 7.024 & 1.275 & 42.121 & 10.925 \\ \cline{2-6}
		%		& n2 & 4.677 & 1.274 & 23.754 & 7.563 \\ \cline{2-6}
		%		& n3   & 5.735 & 0.729 & 27.679 & 0.957 \\ \hline
		%		\hline
		%	\end{tabular}
		%\end{table}
		%
		%\begin{table}[H]
		%	\centering
		%	\begin{tabular}{c|c|ccc||ccc}
		\multicolumn{6}{c}{exponential distribution}\\ \hline \hline
		\multicolumn{2}{c|}{GT} & \multicolumn{2}{c||}{$t=4$} & \multicolumn{2}{c}{$t=8$}\\ \hline
		Cov. Setting &	& $\chi^2$ & WTPS & $\chi^2$ & WTPS    \\ \hline
		\multirow{3}{*}{1} & n1 & 8.416 & 0.431 & 36.706 & 0.968 \\\cline{2-6} 
		& n2 & 9.066 & 1.184 & 37.318 & 1.295 \\ \cline{2-6}
		& n3 & 6.016 & 0.618 & 29.863 & 1.073 \\ \hline\hline
		\multirow{3}{*}{2} & n1  & 8.523 & 0.869 & 36.999 & 1.044 \\\cline{2-6} 
		& n2 & 9.206 & 1.510 & 37.160 & 1.436 \\ \cline{2-6}
		& n3 & 6.445 & 0.293 & 30.130 & 0.925 \\ \hline\hline
		\multirow{3}{*}{3} & n1  & 9.415 & 1.219 & 42.643 & 5.264 \\\cline{2-6} 
		& n2 & 7.638 & 0.946 & 32.131 & 5.851 \\ \cline{2-6}
		& n3 & 6.490 & 0.260 & 30.012 & 0.689 \\ \hline\hline
		%		\multirow{3}{*}{4} & n1  & 9.964 & 2.066 & 51.208 & 14.443 \\\cline{2-6} 
		%		& n2 & 6.859 & 1.121 & 28.327 & 8.702 \\ \cline{2-6}
		%		& n3 & 6.748 & 0.240 & 31.212 & 1.255 \\   \hline
		%		\hline
	\end{tabular}
\end{table}

%\subsection{Large sample behavior}
%
%In addition to the simulation described in the paper we have also analyzed the large sample behavior in covariance setting 4. The results are presented in Figure \ref{Fig.large_Vielfaches}.
%We again note that although the WTS improves greatly with growing number of observations, it still does not keep the pre-assigned level even for 115 observations in each group.
%
%
%\begin{figure}[h]
%	\centering
%	\includegraphics[width=\textwidth]{plot_Vielfaches_200.pdf}
%	\caption{\it Type-I error rates under the interaction hypothesis for the WTS and the WTPS, where sample size was increased by adding $b\veins_3, b=0, 20, \dots, 200$ to the sample size vectors in a balanced (lower panel) and unbalanced (upper panel) design with $t=4$ (left panel) and $t=8$ (right panel) time points under covariance setting 4, i.e., $\vV_i = c_i \cdot \vI_t$ for $\vc = (1, \sqrt{2}, 2),~ i=1, 2, 3$.}
%	\label{Fig.large_Vielfaches}
%\end{figure}

\section{Power}\label{supp:Power}

\begin{figure}[H]
	\centering
	\includegraphics[width=0.7\textwidth]{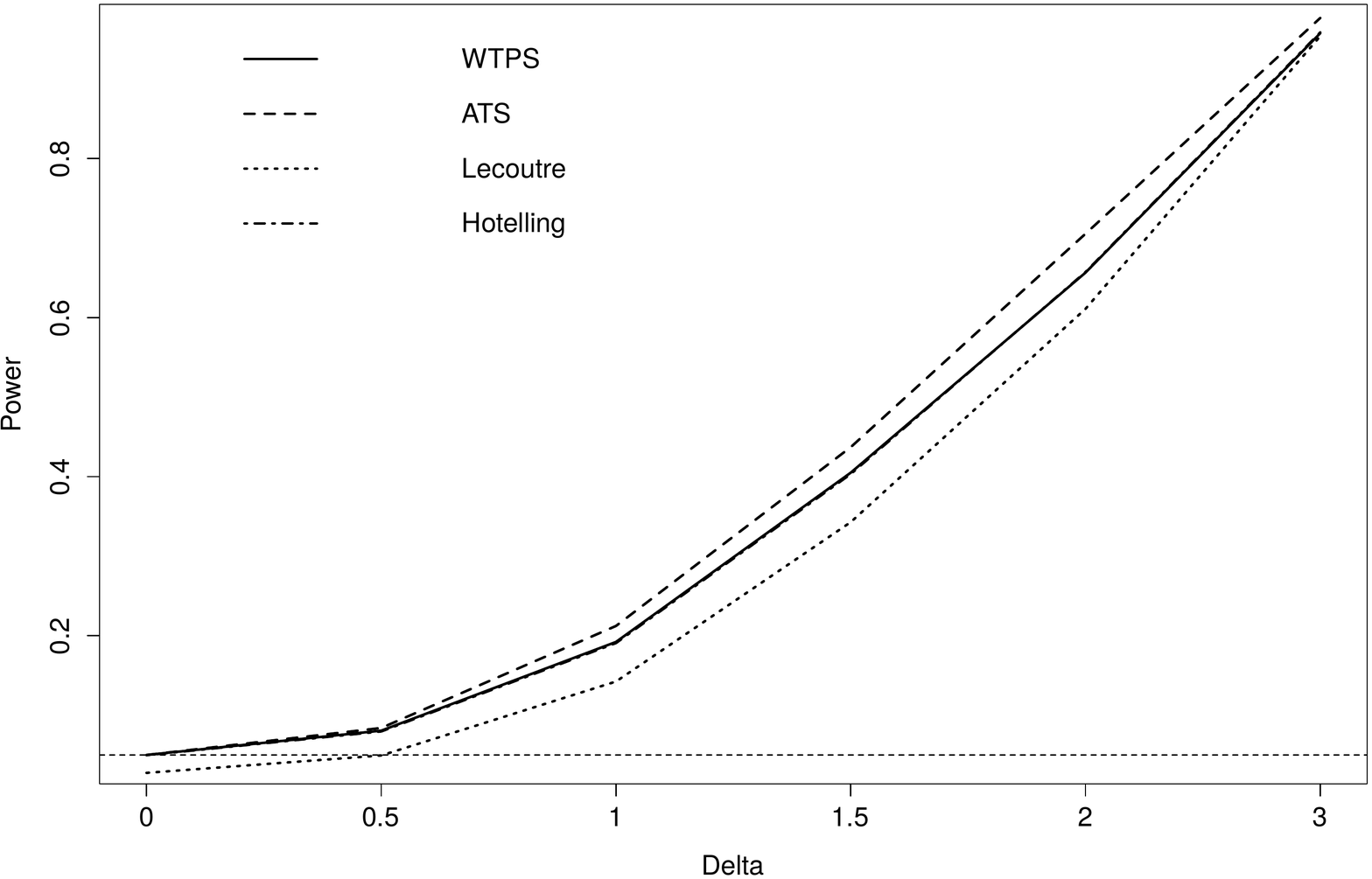}
	\caption{\it Power ($\alpha=0.05$) simulation results of the WTPS, ATS, Lecoutre and Hotelling for normal distribution and 
		$t=4$ repeated measures under a trend alternative $\vmu = (\vmu_1',\vmu_2')'=(\vnull', \delta \vc')'$ with $\delta=\{0, 0.5, 1, 1.5, 2, 3\}$ and $c_s=\frac{s}{t}, s=1, \ldots, t$.}
	\label{fig:Power}
\end{figure}

\begin{figure}[H]
	\centering
	\includegraphics[width=0.7\textwidth]{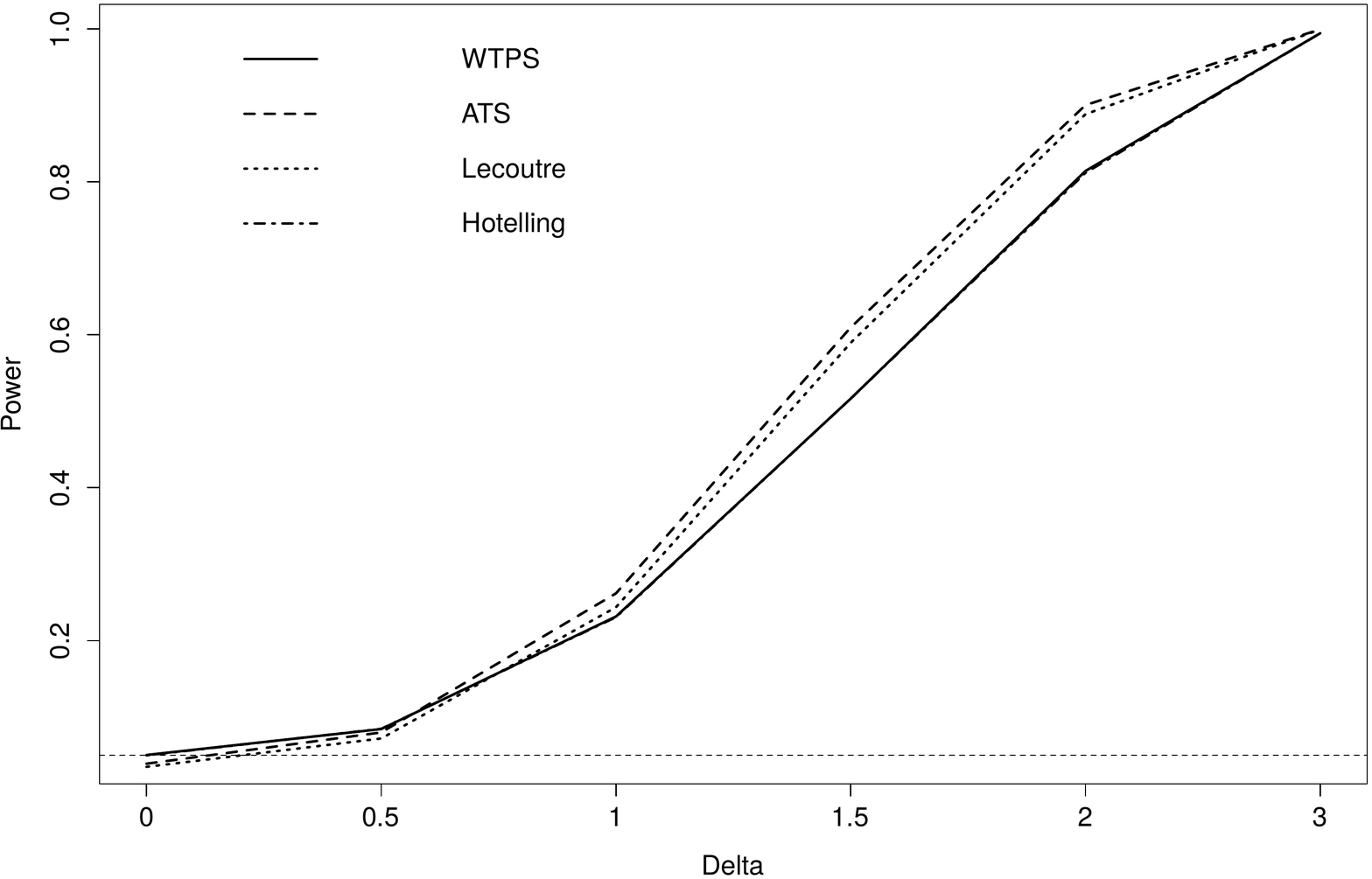}
	\caption{\it Power ($\alpha=0.05$) simulation results of the WTPS, ATS, Lecoutre and Hotelling for normal distribution and $t=8$ 
		repeated measures under a trend alternative $\vmu = (\vmu_1',\vmu_2')'=(\vnull', \delta \vc')'$ with $\delta=\{0, 0.5, 1, 1.5, 2, 3\}$ and $c_s=\frac{s}{t}, s=1, \ldots, t$.}
	\label{fig:Power2}
\end{figure}

\begin{figure}[H]
	\centering
	\includegraphics[width=0.7\textwidth]{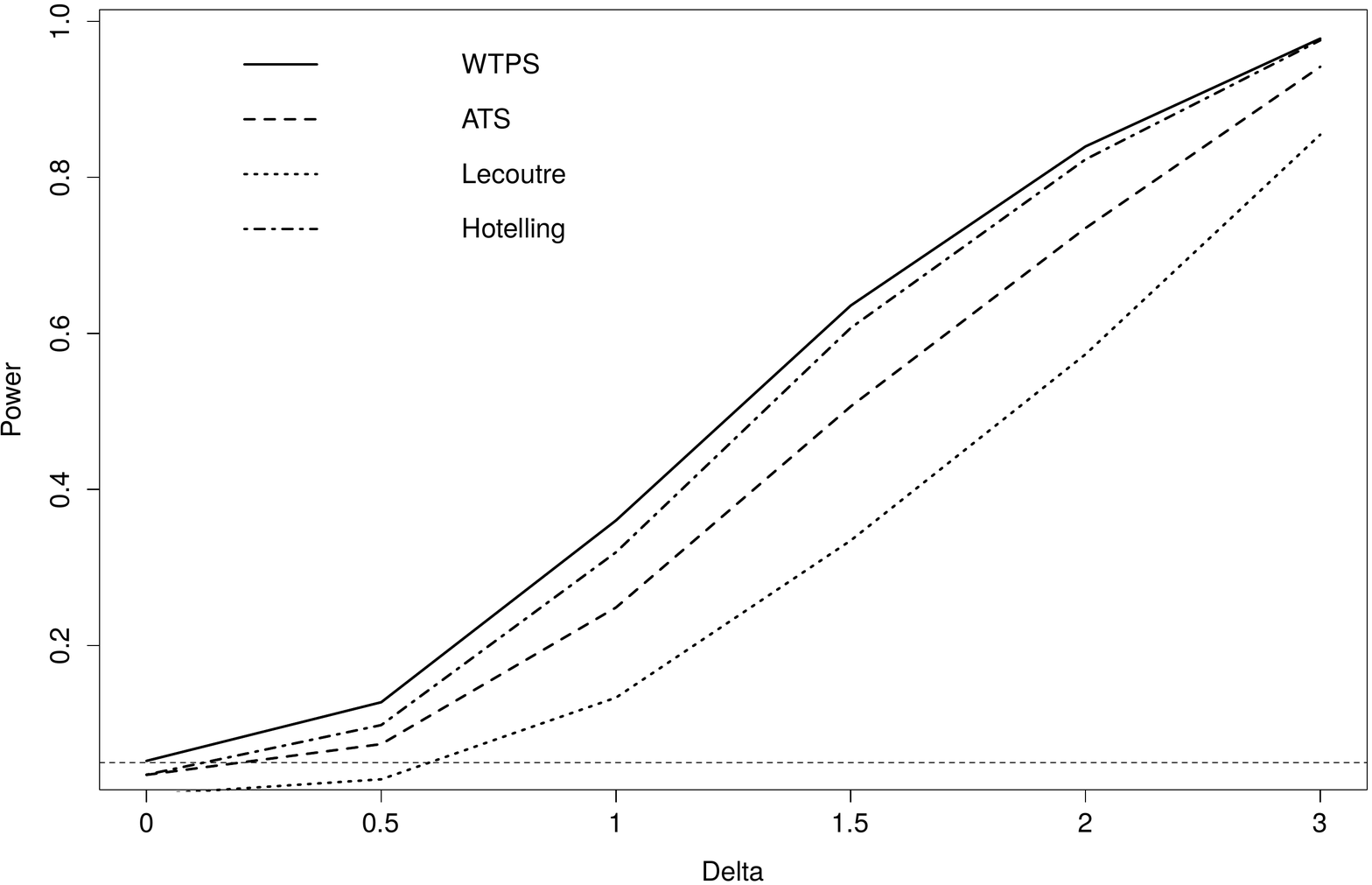}
	\caption{\it Power ($\alpha=0.05$) simulation results of the WTPS, ATS, Lecoutre and Hotelling for log-normal distribution and 
		$t=4$ repeated measures under a trend alternative $\vmu = (\vmu_1',\vmu_2')'=(\vnull', \delta \vc')'$ with $\delta=\{0, 0.5, 1, 1.5, 2, 3\}$ and $c_s=\frac{s}{t}, s=1, \ldots, t$.}
	\label{fig:Power4log}
\end{figure}

\begin{figure}[H]
	\centering
	\includegraphics[width=0.7\textwidth]{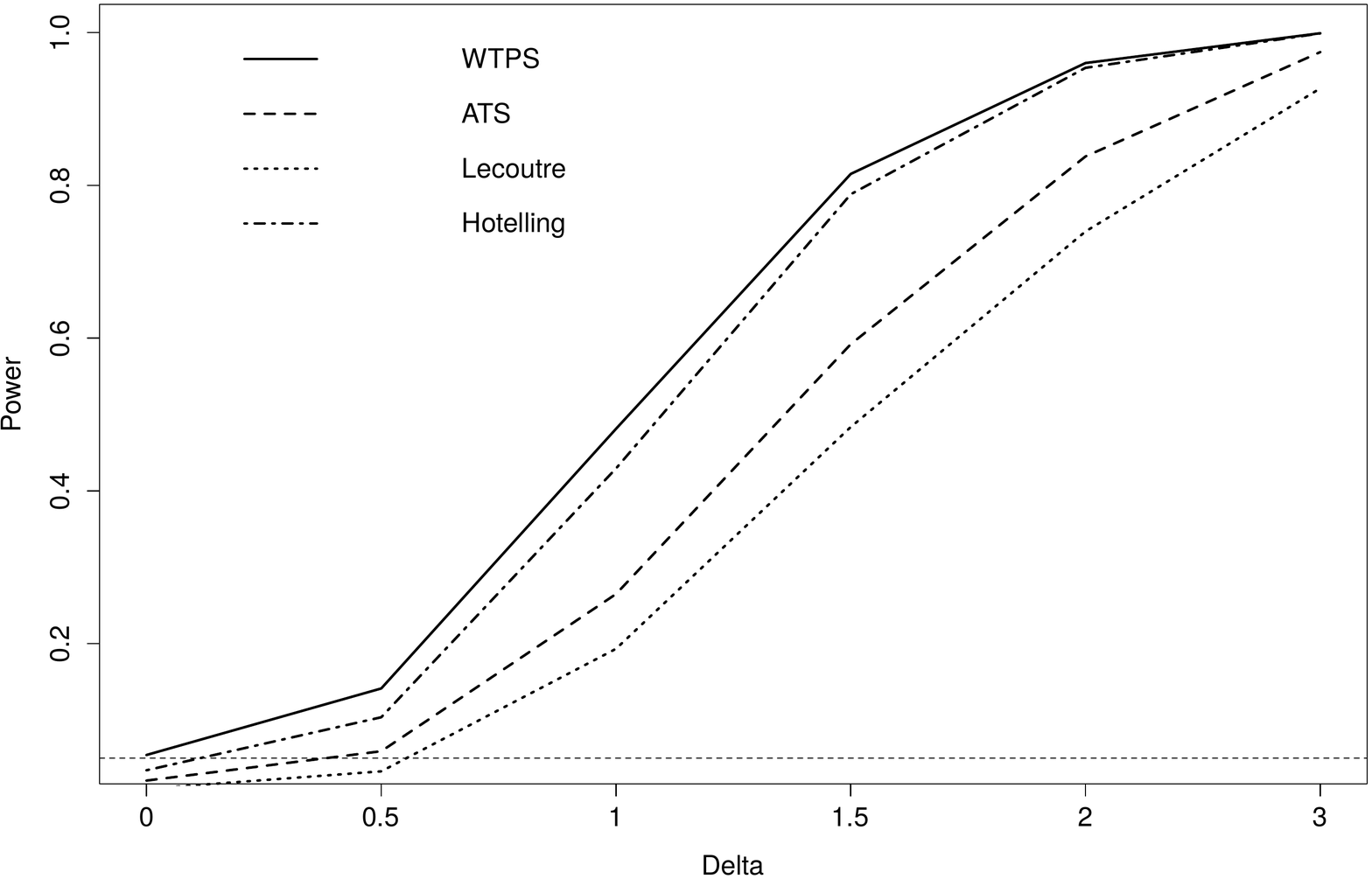}
	\caption{\it Power ($\alpha=0.05$) simulation results of the WTPS, ATS, Lecoutre and Hotelling for log-normal distribution 
		and $t=8$ repeated measures under a trend alternative $\vmu = (\vmu_1',\vmu_2')'=(\vnull', \delta \vc')'$ with $\delta=\{0, 0.5, 1, 1.5, 2, 3\}$ and 
		$c_s=\frac{s}{t}, s=1, \ldots, t$.}
	\label{fig:Power8log}
\end{figure}

We have also conducted several simulations to analyze the power of our method. Since the WTS turned out to test on different $\alpha-$levels (see the simulation results under the null hypothesis), we have excluded it from the analyses. 
%The same holds true for Hotelling's $T^2$ (Hotelling, 1931) and the approximation proposed by Lecoutre (1991) not shown here.\\
We considered a two sample repeated measures design, where we have simulated data as 
$$
\vY_{ik} = (Y_{ik1},\ldots,Y_{ikt})' = \vmu_i +  \vV_i^{1/2} \vep_{ik},
$$
with $\vmu_i=E(\vY_{i1}), i=1,2,$ and $\vV_i \equiv \vI_t$. The i.i.d.~random vectors $\vep_{ik} = (\ep_{ik1}, \dots, \ep_{ikt}), i=1,2,$ were generated from different standardized distributions by
\bqa
\ep_{iks} = \frac{\tilde{\ep}_{iks} - E(\tilde{\ep}_{iks})}{\sqrt{\Var(\tilde{\ep}_{iks})}},
\eqa
where $\tilde{\ep}_{iks}$ denote i.i.d.~normal or log-normal random variables. For the power simulation we have considered a trend alternative, i.e. we set 
$\vmu_2=\vnull$ in the second group and $\vmu_1=\delta \vc = \delta (c_1,\dots,c_t)'$ in the first group, 
where $c_s = \frac{s}{t}, s=1, \ldots, t$ and $\delta \in \{0, 0.5, 1, 1.5, 2, 3\}$. 
We considered a balanced design with 15 individuals per group, hypothesis matrix $\vH = \vP_t (\vI_t\vdots -\vI_t)$ 
\iffalse\[\vH=\vP_t \cdot   \left (\begin{array}{cccccccc}
1 &   0  &  0 &   0 &  -1 &   0 &   0 &   0\\
0  &  1 &   0 &   0 &   0  & -1 &   0  &  0\\
0  &  0  &  1  &  0  &  0  &  0  & -1  &  0\\
0 &   0 &   0 &   1 &   0 &   0  &  0  & -1\\
\end{array} \right)	
\]
\fi
and again simulated both $t=4$ and $t=8$ repeated measures. 
Figures \ref{fig:Power} and \ref{fig:Power2} display the power comparison for the WTPS, the ATS, the approximation described by Lecoutre (1991) as well as Hotelling's $T^2$ (Hotelling, 1931) for normal distribution and $t=4$ and $t=8$ repeated measures, respectively. 
In Figures \ref{fig:Power4log} and \ref{fig:Power8log}, the results for the log-normal distribution are displayed.  
From these figures it appears that the ATS has slightly higher power for normally distributed data. 
%but ---as seen before ---is quite conservative for a large number of repeated measures. 
For log-normally distributed data, the WTPS has larger power than the other methods and it is the only method controlling the type-I error correctly. 
We also note that the approximation by Huynh-Feldt and Lecoutre performs worst for the log-normal distribution.

\section{Analysis of the data example: Comparing the different approaches}\label{supp:DataEx}
We again consider the data example from Section \ref{app} on the oxygen consumption of leukocytes.
First of all, we notice that the empirical covariance matrices of the two groups appear to be quite different. The empirical covariance matrix in the Placebo-group (rounded to three digits) is given as

\[ \left( \begin{array}{cccccc}
0.025 & -0.022 & -0.004 & 0.009 & 0.015 & 0.025\\
-0.022 &  0.092 & -0.005 & -0.001 & -0.024 & -0.035\\
-0.004 & -0.005 & 0.081 & -0.013 & -0.010 & -0.004\\
0.009 & -0.001&  -0.013 & 0.037 & 0.044 & 0.038\\
0.015 & -0.024 &-0.010 & 0.044 & 0.069 & 0.063\\
0.025 & -0.035 & -0.004 & 0.038 & 0.063 & 0.115\\
\end{array} \right)\] 

whereas in the Verum-group we have

\[ \left( \begin{array}{cccccc}
0.043& 0.012& 0.046& 0.033 & 0.014& 0.055\\
0.012& 0.113 & 0.008& 0.009& 0.060& 0.032\\
0.046& 0.008& 0.065& 0.041& 0.005& 0.066\\
0.033& 0.009& 0.041& 0.047& 0.016& 0.059\\
0.014& 0.060& 0.005& 0.016& 0.058& 0.047\\
0.055& 0.032& 0.066& 0.059& 0.047& 0.116
\end{array} \right)\] 
Thus, the assumption of homoscedasticity is not fulfilled in this data example.

The results of the analyses using the different methods are presented in the following table. 
The asymptotic results are again obtained by considering the corresponding $F(\hat{\nu},\infty)$-quantile for the ATS and the $\chi^2_f$-quantile for the WTS.

\begin{table}[h] 
	\centering
	\caption{\it p-values of the analysis of the $O_2$ consumption data.}
	\label{table:data}
	\vspace{0.3cm}
	\begin{tabular}{c|ccc|cccc}
		\hline
		& \multicolumn{3}{|c}{ATS} &  \multicolumn{4}{|c}{WTS}  \\ \hline
		& asymptotic & PBS & NPBS & asymptotic & Permutation & PBS & NPBS \\
		\hline
		A & 0.001   & 0.003 & $<$0.001 &  0.001 & 0.002 & 0.004 & 0.005 \\  
		B & $<$0.001   & 0.001 & 0.001 & $<$0.001 & $<$0.001 & 0.001 & 0.001 \\ 
		T & $<$0.001  & $<$0.001 & $<$0.001 & $<$0.001 & $<$0.001 & $<$0.001 & $<$0.001 \\ 
		AB & 0.110   & 0.130 & 0.140 & 0.110 & 0.118 & 0.125 & 0.136 \\ 
		AT & 0.009   & 0.012 & 0.005 & $<$0.001 & 0.001 & $<$0.001 & $<$0.001 \\
		BT & 0.094   & 0.088 & 0.103 & 0.115 & 0.147 & 0.161 & 0.157 \\ 
		ABT & 0.117   & 0.154 & 0.116 & 0.116 & 0.162 & 0.153 & 0.141 \\ 
		\hline
	\end{tabular}
\end{table}

For this data set, the results are similar for all resampling methods and the asymptotic approaches considered.

\end{document}